%% file: main.tex
\documentclass[acmsmall, screen, nonacm]{acmart}
\AtBeginDocument{%
  }

\setcopyright{acmlicensed}
\copyrightyear{2018}
\acmYear{2018}
\acmDOI{XXXXXXX.XXXXXXX}

\acmJournal{JACM}
\acmVolume{37}
\acmNumber{4}
\acmArticle{111}
\acmMonth{8}

\makeatletter
\AtEndPreamble{%
  \def\@acmdefinitionindent{\z@}%
}
\makeatother

\usepackage{booktabs}   
\usepackage{multirow}   
\usepackage{subcaption} 
\usepackage{listings}
\usepackage{parcolumns}
\usepackage{placeins}
\usepackage{multicol} 
\usepackage[utf8]{inputenc}
\usepackage{mathpartir}

\usepackage[font=small,skip=0pt]{caption}
\usepackage{tikz-cd}
\usepackage{graphicx}

\usepackage{chor}
\usepackage{symbols}

\newcommand{\mech}{Mech\xspace}
\newcommand{\async}{Nyarlathotep\xspace}

\definecolor{kirby}{RGB}{215,72,148}
\definecolor{kirby-blue}{RGB}{68,102,191}
\definecolor{kirby-green}{RGB}{35,135,37}

\newcommand{\mathshade}[1]{\mathpalette\mathshadeaux{#1}}
\newcommand{\mathshadeaux}[2]{%
  \begingroup
  \setlength{\fboxsep}{2pt}%
  \colorbox{black!10}{$\mathsurround=0pt#1\mathstrut#2$}%
  \endgroup
}

\usepackage{cleveref} 
\crefname{example}{example}{examples}
\Crefname{example}{Example}{Examples}
\crefname{proposition}{proposition}{propositions}
\Crefname{proposition}{Proposition}{Propositions}

\usepackage{enumitem}

\makeatletter
\AtBeginDocument{%
  \apptocmd{\ps@standardpagestyle}{%
    \fancyhead[LE,RO]{}%
    \fancyfoot[C]{\footnotesize\thepage}%
  }{}{}%
  \apptocmd{\ps@firstpagestyle}{%
    \fancyfoot[C]{\footnotesize\thepage}%
  }{}{}%
  \pagestyle{standardpagestyle}%
}
\makeatother

\begin{document}

\title{A Principled Approach to Concurrency, Nondeterminism, and Asynchrony in Choreographies}

\author{Xueying Qin}
\email{xyqin@imada.sdu.dk}
\orcid{0000-0003-4825-2023}
\affiliation{%
  \institution{University of Southern Denmark}
  \city{Odense}
  \country{Denmark}
}

\author{Marco Peressotti}
\email{peressotti@imada.sdu.dk}
\orcid{0000-0002-0243-0480}
\affiliation{%
  \institution{University of Southern Denmark}
  \city{Odense}
  \country{Denmark}
}

\author{Fabrizio Montesi}
\email{fmontesi@imada.sdu.dk}
\orcid{0000-0003-4666-901X}
\affiliation{%
  \institution{University of Southern Denmark}
  \city{Odense}
  \country{Denmark}
}


\begin{abstract}
Choreographic programming (CP) describes a distributed system as a global program and compiles it into communicating network programs via endpoint projection (EPP). Asynchrony captures the independent progress of communicating processes, while nondeterminism captures the range of behaviours that can arise from their interactions. However, the integration of both features into a theory of CP is absent.

We present \async{}, the first asynchronous and nondeterministic CP framework.
We define its operational semantics of choreographies and networks using labelled transition systems.
Besides asynchrony and choreographic nondeterminism, \async{} supports
expressive features including general branching, out-of-order execution, and
general recursion with parameterised procedures that processes can enter
independently. We prove EPP sound and complete, establishing strong behavioural
equivalence between the semantics of choreographies and their
projected networks. We prove that projected networks are
communication-safe and deadlock-free. We mechanised all our results in Lean~4.
\end{abstract}




\maketitle
\input{sections/introduction}
\input{sections/background}
\input{sections/choreography}
\input{sections/process-network}
\input{sections/nondeterminism-async}
\input{sections/epp}
\input{sections/epp-correct}
\input{sections/comm-safe-deadlock-free}
\input{sections/related-work}
\input{sections/conclusion-future-work}
\bibliographystyle{ACM-Reference-Format}
\bibliography{references}

\appendix
\newpage
\input{appendix/pn}
\input{appendix/store-and-subst}
\input{appendix/wf}
\input{appendix/merge}
\input{appendix/coh}
\input{appendix/algebraic-properties}
\input{appendix/local-semantics}
\input{appendix/queue-transition}
\input{appendix/finfun}
\input{appendix/process-ops}
\input{appendix/semantic-properties}
\input{appendix/misc}
\end{document}

%% file: sections/introduction.tex
\section{Introduction}
\label{sec:introduction}

Traditionally, concurrent and distributed systems are programmed from a \emph{local perspective}.
Developers write programs for individual endpoints, and the overall behaviour of the system emerges from interactions amongst those endpoints.
With such an approach, concurrent and distributed systems are prone to critical errors including deadlocks and mismatched communications, and the overall system behaviours are hard to reason about.

\emph{Choreographic programming} (CP) is an emerging correct-by-construction programming paradigm, which addresses these issues by enabling developers to program concurrent and distributed systems from a \emph{global perspective}~\citep{M13:phd,M23}.
A \emph{choreography} is a global program that specifies the intended computation and the interactions among endpoints, which are \emph{processes}.
\emph{Endpoint projection (EPP)} automatically compiles a choreography into a \emph{network} of endpoint programs that follow the behaviours specified by the choreography.
The correctness of EPP establishes \emph{behavioural equivalence} between choreographies and their projected programs. Consequently, the safety and liveness properties characterised by a choreography are preserved in networks.
In practice, there are many stand-alone and library-level implementations of CP, such as Choral~\cite{GMP24}, Klor~\cite{LJ24}, Chorex~\cite{WG25}, HasChor~\cite{SKK23}, MultiChor~\cite{BKJSKN25}, and UniChorn~\cite{C25}.

Theories of CP have been rapidly developed over the past few years, and some have been accompanied by substantial mechanisation efforts, opening up opportunities to build certified compiler frameworks for programming concurrent and distributed systems. However, an asynchronous CP framework with choreographic nondeterminism is still missing. 

Asynchrony and nondeterminism are each difficult to model. The textbook semantics for nondeterministic choreographic choice is wrong~\cite{M23Errata,QPM26}.
In the multiparty session types (MPST) community, researchers have found flaws in subject-reduction proofs~\cite{CGY21,SY19,TBC25b} and shown that a previously proposed asynchronous subtyping relation is undecidable~\cite{BCZ17}.
Work on mixed choice highlights the difficulty of coordinating participants in the presence of nondeterminism, under both synchronous and asynchronous communication~\cite{PY24a,BHVT26}.

However, an asynchronous nondeterministic CP framework is necessary for expressing complex system behaviours. In \cref{ex:reusable-lock}, we use a \emph{recursive choreographic procedure definition} to model a reusable lock. With asynchrony and nondeterminism, two processes $\pid{p_1}$ and $\pid{p_2}$ repeatedly compete to update a shared inventory at $\pid{s}$, which manages the lock.
$\pid{s}$ replies with $\lbl{ok}$ to the first request it receives, \emph{regardless} of the order in which the requests were sent.
After sending $\lbl{ok}$, $\pid{s}$ can take a local snapshot of the inventory while the reply is still queued for the winner.
The winner works with $\pid{s}$ in the \emph{procedure} $W$, where $\pid{s}$ can use the snapshot to restore the inventory if an update fails. The completion of $W$ serves as an implicit release.
The notation $C_1 \choicep{s} C_2$ denotes a nondeterministic \emph{choreographic choice} between $C_1$ and $C_2$, which $\pid{s}$ resolves by taking a step in one of the branches.
\begin{example}[A Deadlock-Free Reusable Lock]\label[example]{ex:reusable-lock}
\begin{spreadlines}{0pt}
\begin{align*}
X(\pid{s},\pid{p_1},\pid{p_2}) &= (\csel{p_1}{s}{req}\seq \csel{s}{p_1}{ok}\seq \assign{s}{\vname{backup}}{\fname{snapshot}(\vname{inventory})}\seq\\
&\quad\;\; W(\pid{p_1},\pid{s})\seq \csel{p_2}{s}{req}\seq \csel{s}{p_2}{ko}\\
&\choicep{s}\csel{p_2}{s}{req}\seq \csel{s}{p_2}{ok}\seq \assign{s}{\vname{backup}}{\fname{snapshot}(\vname{inventory})}\seq\\
&\quad\;\; W(\pid{p_2},\pid{s})\seq \csel{p_1}{s}{req}\seq \csel{s}{p_1}{ko})\seq X(\pid{s},\pid{p_1},\pid{p_2})
\end{align*}
\end{spreadlines}
\end{example}

Endpoint projection produces \emph{process procedure definitions} for the reusable lock, shown in \cref{ex:net-fcfs-rec}, each specifying the local behaviour of one of the participating processes. Each process can enter its procedure call independently.
Repeatedly, $\pid{p_1}$ and $\pid{p_2}$ send $\lbl{req}$ to $\pid{s}$ ($\pselg{s}{req}$), each participating in $W$ only upon receiving $\lbl{ok}$. Branches labelled by $\lbl{ok}$ and $\lbl{ko}$ are \emph{merged} into a \emph{branching term} that waits to receive a selection label from $\pid{s}$.
The first request that $\pid{s}$ receives resolves the \emph{nondeterministic choice} ($\choice$). If both requests are queued, either can win.
After sending $\lbl{ok}$, $\pid{s}$ can take a local snapshot of the inventory while the reply is still queued. It then executes its part of $W$ before receiving and rejecting the other request with $\lbl{ko}$ and repeating the procedure.
The $(\fatsemi)$ notation here is the \emph{sequential composition} of process terms, formally defined in \cref{sec:proc:semantics:seqcomp}.
\begin{example}[A Reusable Lock as Process Procedure Definitions]\label[example]{ex:net-fcfs-rec}
The two process procedure definitions for clients $\pid{p_1}$ and $\pid{p_2}$ are:
$X_2(\pid{s},\pid{p_2}) = \pselg{s}{req} \seq \pid{s}\br\{\lbl{ok} : W_1(s)\sep \lbl{ko} : \pnil \} \fatsemi (X_2(\pid{s},\pid{p_2}))$,
and
$X_3(\pid{s},\pid{p_1}) = \pselg{s}{req} \seq \pid{s}\br\{\lbl{ko} : \pnil \sep \lbl{ok} : W_1(s)\} \fatsemi (X_3(\pid{s},\pid{p_1}))$.
The definition for the server $\pid s$ is:
\begin{spreadlines}{0pt}
\begin{align*}
X_1(\pid{p_1},\pid{p_2}) &= (\pid{p_1}\br\{\lbl{req}: \pselg{p_1}{ok}\seq \vname{backup}\metaDef \fname{snapshot}(\vname{inventory})\seq\\
&\quad\;\; W_2(\pid{p_1})\seq \pid{p_2}\br\{\lbl{req}: \pselg{p_2}{ko}\}\}\\
&\choice \pid{p_2}\br\{\lbl{req}: \pselg{p_2}{ok}\seq \vname{backup}\metaDef \fname{snapshot}(\vname{inventory})\seq\\
&\quad\;\; W_2(\pid{p_2})\seq \pid{p_1}\br\{\lbl{req}: \pselg{p_1}{ko}\}\})\seq X_1(\pid{p_1},\pid{p_2})
\end{align*}
\end{spreadlines}
\end{example}

We have studied existing theories of CP that could serve as a starting point to developing a theory expressive enough to express examples like \cref{ex:reusable-lock}. We see two possible expressive and machine-checked candidates, with different approaches to concurrency. The state-of-the-art functional choreographic language $\lambda_{\mathrm{QC}}$~\cite{SHC25} extends Pirouette~\cite{HG22}, supporting features such as process polymorphism and higher-order functions. However, it only supports synchronous communication. On top of that, invoking a function application requires all processes to synchronise, even those uninvolved in the call. This design not only conflicts with the concurrent semantics of distributed systems but also poses substantial obstacles to adding choreographic nondeterminism. Introducing $\choicep{s}$ under $\lambda_{\mathrm{QC}}$'s global application rule leaves the example lock unprojectable: $\pid p_1$ and $\pid p_2$ cannot enter the procedure calls independently, a step for the losing process to joining the procedure need to be added, which causes two branches unmergeable. Following the textbook approach~\cite{M23}, \mech~\cite{QPM26} does not require global synchronisation and correctly models the semantics of nondeterministic choreographic choice. But its synchronous communication produces cumbersome nondeterministic semantics and limits the behaviours it can express (cf. \cref{sec:nondeterminism-async}). \mech's approach to concurrency and algebraic characterisations for choreographies still make it a good starting point for extending CP.

In this paper we present \async, a semantic foundation for asynchronous and nondeterministic choreographies, which also features out-of-order execution, general branching, and general recursion parameterised over processes that allows processes to enter calls independently.
Our paper makes the following contributions:
\begin{itemize}[nosep]
\item We present the first formalism for asynchronous and nondeterministic choreographies. We design new operational semantics for choreographies and networks to capture the interplay amongst asynchrony, nondeterminism, and out-of-order execution.
\item \async's operational semantics has a new \emph{locality property}: unlike in \mech, where synchronous transitions can involve multiple processes, in \async each transition concerns only one active process, both in choreographies and networks. This property enables more structured and compositional reasoning about the correctness of EPP.
\item We prove the correctness of EPP, which yields networks behaviourally equivalent to their source choreographies and guarantees communication safety and deadlock freedom.
\item We mechanise all results in Lean~4, providing a machine-checked formal framework that enables the future development of a certified compiler for CP. Benefitting from the locality property, the size of the mechanisation of \async is only around half of \mech's.
\end{itemize}

%% file: sections/background.tex
\section{Background}
\label{sec:background}
The design of \async builds on \mech's account of concurrency in CP. We briefly introduce core reusable features of \mech and give a high-level overview of CP.
\subsection{Mech}
\label{sec:background:mech}
\mech is a first-order choreographic calculus mechanised in Lean~4.
It models synchronous communication and supports multi-way branching, general recursion with process parameters and independent procedure entry, and nondeterministic choreographic choice.
\mech abstracts over local computations by taking an evaluation relation
for local expressions as a parameter. This allows the choreographic
semantics to describe how processes interact independently of how they
compute their local results.
We use \mech's invoicing choreography \cref{ex:CinvC} to give an overview of the design of a choreographic programming language.
\begin{example}[An Invoicing Choreography]\label[example]{ex:CinvC}
The shop $\pid{p}$ decides whether to request payment, issue a reimbursement, or skip invoicing. The guards are evaluated at $\pid{p}$. Since the customer $\pid{q}$ has a different role in each branch, $\pid{p}$ sends it a selection label before proceeding.
\begin{spreadlines}{0pt}
\begin{align*}
\cn{inv} \defeq &\condif \pe{p}{\fname{pay()}} \condthen \csel{p}{q}{pay}\seq\ccom{\pid{p}}{\fname{invoice()}}{\pid{q}}{x} \seq \cnil\\
&\condelse \condif \pe{p}{\fname{reimburse()}} \condthen \csel{p}{q}{reimburse}\seq \ccom{\pid{q}}{\fname{coords()}}{\pid{p}}{c} \seq \ccom{\pid{p}}{\fname{invoice(c)}}{\pid{q}}{x} \seq \cnil\\
&\condelse \csel{p}{q}{skip}\seq\cnil
\end{align*}
\end{spreadlines}
\end{example}

\mech defines operational semantics for both choreographies and
endpoint networks. At the choreographic level, actions on disjoint
processes can execute out of order. An assignment at a third process,
for instance, can proceed independently of the exchange of bank
details. The exchange itself is synchronous. The send at $\pid{q}$
and the matching receive at $\pid{p}$ occur in a single transition.

EPP produces the network in \cref{ex:EPPinvC}, where each endpoint runs
the following program.
\begin{example}[EPP of the Invoicing Choreography]\label[example]{ex:EPPinvC}
The shop $\pid p$, as a guard process, retains the two evaluations of guard expressions. The customer $\pid q$'s three possible behaviours are \emph{merged} into one branching term that waits for a label from the shop. We describe this merge operation in \cref{sec:background:merge}.
\begin{spreadlines}{0pt}
\begin{align*}
\left.
\begin{aligned}[c]
\epp{\cn{inv}} = \
\end{aligned}
\begin{aligned}[c]
\atomicn{p}{&\condif\fname{pay()}\condthen \\&\quad\pselg{q}{pay} \seq\psend{q}{\fname{invoice()}}\seq\pnil\\ &\condelse \condif \fname{reimburse()} \condthen \\&\quad\pselg{q}{reimburse}\seq\precv{q}{c}\seq\psend{q}{\fname{invoice(c)}}\seq\pnil \\ &\condelse \pselg{q}{skip}\seq\pnil }
\end{aligned}
\; \middle|\;
\begin{aligned}[c]
\atomicn{q}{\pid{p}\br\{&\lbl{pay}: \precv{p}{x}\seq\pnil\sep\\&\lbl{reimburse}: \psend{p}{\fname{coords()}} \seq \precv{p}{x} \seq \pnil \sep\\&\lbl{skip}: \pnil \}}
\end{aligned}
\right.
\end{align*}
\end{spreadlines}
\end{example}

\subsection{Select and Merge: the Algebraic Approach to Handling Knowledge of Choice (KoC)}
\label{sec:background:merge}

The \emph{select-and-merge} approach handles \emph{knowledge of choice}
(KoC) in choreographic programming~\citep{M13:phd,M23}.
As shown in \cref{ex:EPPinvC}, a selection communicates the outcome
of a choice as a label, and merge combines the projected alternatives
into a single endpoint program.

We adopt \mech's formulation of the selection and
\emph{full merge}~\citep{DYBH12} approach. EPP preserves a conditional
at the process that evaluates its guard and projects a choreographic
choice to a nondeterministic choice at the choice-resolving process.
At every other process, EPP projects the alternatives separately and
combines them using the merge operation $(\sqcup)$.

\mech defines merge on optional process terms
($\opt{Process}$) and optional instructions ($\opt{PInstr}$) by mutual recursion.
We formally introduce the process language in \cref{sec:proc-net:syntax}.
A successful merge of process terms returns $\some{P}$ for the
combined term $P$. Incompatible terms and failed inputs yield $\none$.
Two branching terms can merge when they expect a label from the same
sender. The result offers the union of their labels. A label offered
by only one term keeps its continuation, while a label offered by both
requires their continuations to merge recursively.

Merge is associative, commutative, and idempotent, making
$(\opt{Process},\sqcup)$ a \emph{join-semilattice}. It introduces a \emph{partial order}, which is the \emph{more branches} relation: $\opterm{P} \mbrs \opterm{Q}$ if and only if $\opterm{P}\sqcup\opterm{Q}=\opterm{P}$. Note that $\opterm{P}$ denotes a process term that can either be $\some{P}$ or $\none$.
$\mbrs$ is a \emph{simulation relation}, meaning that $P$ offers every label that $Q$ offers at the corresponding branching terms and may offer additional
labels. The order enables us to compare a residual network that still offers several alternatives with the projection of a choreography that has already resolved a choice.
For the formal definitions of merge and the more branches relation, and for their properties, see \cref{app:merge}.

\subsection{Formal Preliminaries}
\label{sec:formal:pre}
We introduce formal preliminaries we adopt from \mech in the development of \async.
\paragraph{Parameters}
\label{sec:parameters}
To design a local language agnostic CP framework, our models do not fix internal details of local computations at processes.
Specifically, we require the following parameters:
\begin{itemize}[nosep]
\item A set of process names ($\setof{PName}$) used to identify processes, ranged over by $\pid p$.
\item A set of values ($\setof{Val}$) that processes can compute, ranged over by $v$.
\item A set of variables ($\setof{Var}$) that processes use to store values, ranged over by $x$.
\item A set of function names ($\setof{F}$) used to identify local functions that processes can invoke.
\item A set of selection labels ($\setof{SLabel}$) used by processes to offer alternative behaviours.
\item A set of procedure names ($\setof{ProcName}$) used to identify recursive procedures in programs.
\end{itemize}

\paragraph{Function Support}
\label{sec:finfun}
Given a function $f$ whose codomain has a zero element, we write $\support(f)$ for its support, defined by: $ \support(f) \triangleq \{ a \in A \mid f(a) \neq 0 \}$.

\paragraph{Expressions, Stores, and Evaluation of Expressions}
\label{chor:semantics:stores}
The local language expressions are of the form:
\[\mathrm{Expression\;(\setof{Expr})} \quad e \metaDeff \ v \cmid x \cmid f(\vec{e})\]
A global store $\cstore$ is a function mapping process names to \emph{local stores}, representing the memory state of the system. A local store $\sigma$ is a function mapping variables to values, representing the local memory of a process. We define them as follows:
\begin{align*}
\begin{aligned}[c]
    \mathrm{Global\, Store\, (\setof{GStore})}\quad \cstore : \setof{PName} \to \setof{LStore}
\end{aligned}
\quad\quad\quad\quad
\begin{aligned}[c]
    \mathrm{Local\, Store\, (\setof{LStore})} \quad\sigma : \setof{Var} \to \setof{Val}
\end{aligned}
\end{align*}
\\[-3.5ex]
We write $\sigma[x \mapsto v]$ for the \emph{update} of $x$ to $v$ in a local store, and $\cstore[\pid{p}.x \mapsto v]$ for the corresponding update in the local store of $\pid{p}$ (cf. definition~\ref{app:def:store-update}). The \emph{evaluation} relation $\sigma \vdash e \downarrow v$ states that $e$ evaluates to $v$ under $\sigma$. We assume only that $\downarrow$ is total; evaluation can be \emph{nondeterministic}.

%% file: sections/choreography.tex
\section{Choreographies}
\label{sec:chor}
We introduce the formal syntax and semantics of choreographies in \async. Throughout the paper, we shade the components in our formalism that model asynchrony and choreographic nondeterminism. While designing new semantics for both choreographies and networks, we add only two runtime terms and retain the surface syntax of \mech. This makes \async an ergonomic extension of \mech modelling richer behaviours with the same program encodings.
\subsection{Syntax}
\label{chor:syntax}
\begin{figure}
\vspace{-1.0em}
\begin{spreadlines}{0pt}
\begin{align*}
    \text{Choreographic Procedure Definitions ($\mathfrak{C}$)} && \mathscr{C} \metaDeff &\ \{X_i(\pids{p_i}) = C_i\}_{i \in \idx{I}}\\
    \mathrm{Choreography\;(\setof{Chor})} &&
    C \metaDeff&\ I \seq C \cmid \cnil\\
    \mathrm{Choreographic\;Instruction\;(\setof{Instr})} &&
    I \metaDeff&\ \pid{p}.x \metaDef e \cmid \gencom \cmid \gensel \\ && \cmid &\ \mathshade{\genrtcom} \cmid \mathshade{\genrtsel} \cmid \condif\pid{p}.e\condthen C_1\condelse C_2 \\ && \cmid &\ C_1 \choice_\pid{p} C_2 \cmid X(\pids{p}) \cmid \pid{q} : X(\pids{p}).C
\end{align*}
\end{spreadlines}
\caption{\label{chor:syntax:fig}Syntax of Choreographies in \async}
\vspace{-1.0em}
\end{figure}
\Cref{chor:syntax:fig} shows the formal syntax of choreographies in \async, where we mutually define choreographies ($C$) and their instructions ($I$).
A choreography is either a sequence of instructions, or a terminated choreography $\cnil$.
For readability, we often omit trailing $\cnil$s in examples.
The two new runtime instructions $\genrtcom$ and $\genrtsel$ represent a pending receive at $\pid{q}$ after $\pid{p}$ sends a value or a selection label, respectively.
We define standard functions $\pn(C)$ and $\pn(I)$ for gathering the (active) process names in choreographies and instructions. \Cref{chor:pn:async-choice} shows selected definitions. These functions are particularly important for modelling out-of-order execution in the semantics.

\subsection{Semantics}
\label{chor:semantics}
We define the operational semantics of choreographies as a \emph{labelled transition system (LTS)} whose states are configurations of the form $\langle C, \cstore, \mst, \mathscr{C} \rangle$, where $C$ is a choreography, $\cstore$ is a \emph{global store} mapping process names to their \emph{local stores}, $\mst$ is a \emph{messaging state}, and $\mathscr{C}$ is a set of \emph{choreographic procedure definitions}.
A transition takes the form of $\langle C, \cstore, \mst, \mathscr{C}\rangle \xrightarrow[]{\mu} \langle C', \cstore', \mst', \mathscr{C}\rangle$, where $\mu$ is a transition label. We first introduce some important notations and definitions used in our semantics.

\subsubsection{Transition Labels}
\label{chor:semantics:tl}
\begin{figure}[b]
\vspace{-1.0em}
\begin{spreadlines}{0pt}
\begin{alignat*}{3}
\text{Transition Label} \quad && \mu \ \metaDeff& \ \tlassign{p} \cmid \mathshade{\tlcoms{p}{v}{q}} \cmid \mathshade{\tlcomr{p}{v}{q}} \cmid \mathshade{\tlsels{p}{q}{l}} \cmid \mathshade{\tlselr{p}{q}{l}} \\ && \cmid &\ \tlthen{p} \cmid \tlelse{p} \cmid \tlcall{p} \cmid \tlskip{p} \cmid \mathshade{\tlleft{p}{\mu}} \cmid \mathshade{\tlright{p}{\mu}}
\end{alignat*}
\end{spreadlines}
\caption{Transition Labels in \async}
\label{chor:semantics:transition-labels}
\vspace{-1.0em}
\end{figure}
A transition label $\mu$ for the LTS in \mech is defined as a pair $(\rho, \zeta)$, where $\rho$ is a pre-transition label indicating the action performed and $\zeta$ maps process names to lists of \emph{choice labels}, $\mathit{left}$ or $\mathit{right}$, indicating the chosen branches for a choice-resolving process and carrying the communication synchronisation obligation in choreographic nondeterminism.

Since \async models asynchronous communications, transition labels no longer need to carry the synchronisation obligation, yielding a straightforward design which takes a list structure.

There are two major changes to model the asynchronous semantics. First, sending and receiving actions are treated as two separate steps, reflected in $\tlcoms{p}{v}{q}$, $\tlcomr{p}{v}{q}$, $\tlsels{p}{q}{l}$, and $\tlselr{p}{q}{l}$. Second, transition labels for modelling nondeterministic behaviours now become cleaner: it states the choice-resolving behaviour without carrying the synchronisation obligations. We discuss the interaction between asynchrony/synchrony and nondeterminism in \cref{sec:nondeterminism-async}.
We also define the function $\pn$ for gathering active process names in a transition label, shown in~\cref{chor:pn:async-choice}

\begin{figure}
\newcommand{\pnindent}[2]{\hspace{\widthof{$\pn(#1)$}-\widthof{$\pn(#2)$}}}
\[
\begin{gathered}
\text{(Process names $\pn$ in choreographies and instructions)} \\
\textstyle
  \pn(\gencom) \triangleq \{\pid{p},\pid{q}\} \quad
  \mathshade{\pn(\genrtcom) \triangleq \{\pid{q}\}} \quad
  \pn(\gensel) \triangleq \{\pid{p},\pid{q}\} \quad
  \mathshade{\pn(\genrtsel) \triangleq \{\pid{q}\}} \\[0.8ex]
\text{(Process names $\pn$ in transition labels)} \\
\begin{array}{@{}l@{\qquad}l@{\qquad}l@{}}
  \pnindent{\tlsels{p}{q}{l}}{\tlcoms{p}{v}{q}}
  \mathshade{\pn(\tlcoms{p}{v}{q}) \triangleq \{\pid{p}\}}
  & \pnindent{\tlselr{p}{q}{l}}{\tlcomr{p}{v}{q}}
    \mathshade{\pn(\tlcomr{p}{v}{q}) \triangleq \{\pid{q}\}}
  & \pnindent{\tlright{p}{\mu}}{\tlleft{p}{\mu}}
    \mathshade{\pn(\tlleft{p}{\mu}) \triangleq \{\pid{p}\} \cup \pn(\mu)} \\[2pt]
  \mathshade{\pn(\tlsels{p}{q}{l}) \triangleq \{\pid{p}\}}
  & \mathshade{\pn(\tlselr{p}{q}{l}) \triangleq \{\pid{q}\}}
  & \mathshade{\pn(\tlright{p}{\mu}) \triangleq \{\pid{p}\} \cup \pn(\mu)}
\end{array}
\end{gathered}
\]
\caption{Selected Definitions of $\pn$ (Full definitions are in definitions~\ref{chor:syntax:pn} and \ref{chor:semantics:pntl})}
\label{chor:pn:async-choice}
\end{figure}

\subsubsection{Monoidal Construction of Sequential Composition}
In \cref{chor:semantics:seqcomp}, we define a \emph{semantic operator}, the \emph{sequential composition operator} ($\fatsemi$), for handling the continuation of choreographies in transitions.
\begin{figure}[b]
\vspace{-1.0em}
\begin{align*}
\begin{aligned}[c]
    \cnil \fatsemi C &\triangleq C\\[-0.5ex]
    (I \seq C') \fatsemi C &\triangleq (I \fatsemi C) \seq (C' \fatsemi C)
\end{aligned}
\quad\quad\quad\quad
\begin{aligned}[c]
    I \fatsemi C &\triangleq 
    \begin{cases}
    \ \pid{q} : X(\pids{p}).(C' \fatsemi C) &\text{if $I = \pid{q} : X(\pids{p}).C' $}\\[-0.5ex]
    \ I &\text{otherwise}
    \end{cases}
\end{aligned}
\end{align*}
\caption{Sequential Composition of Choreographies}
\label{chor:semantics:seqcomp}
\vspace{-1.0em}
\end{figure}
The set of choreographies together with the sequential composition operator $\fatsemi$ forms a \emph{monoid}, where $\cnil$ is the \emph{identity element} and $\fatsemi$ is the \emph{associative operator} (see proposition~\ref{app:chor:seqcomp:prop}).

\subsubsection{Messaging State}
\label{chor:semantics:messaging-state}
The messaging state $\mst$ in \async records messages that have been sent and are awaiting receipt.
A message $\msg$ is either a value $v$ or a selection label $\albl$, and $[\setof{M}]$ denotes lists of messages.
\begin{align*}
\text{Message $(\setof{M})$} \quad m \ \metaDeff \ v \cmid \albl \quad\quad\quad
\text{Messaging State} \quad \mst \ : \ \setof{PName} \to \setof{PName} \to [\setof{M}]
\end{align*}
For each sender $\pid{p}$ and receiver $\pid{q}$, $\mst(\pid{q}, \pid{p})$ is a first-in, first-out (FIFO) queue of messages from $\pid{p}$ to $\pid{q}$. We write $ \mst[(\pid{q},\pid{p}) \mapsto \vec{m}]$ for the \emph{update} that sets the queue from $\pid{p}$ to $\pid{q}$ to $\vec{m}$.

\subsubsection{LTS for Choreographies}
\label{chor:semantics:lts}
\Cref{chor:semantics:lts:fig-1} gives the rules for the asynchronous communications and choreographic choice. A value
communication $\gencom$ takes two transitions. In \textsc{send-val},
$\pid{p}$ evaluates $e$ in its local store. The resulting value
$v$ is then appended to the message queue for the communication from $\pid{p}$ to $\pid{q}$ in the messaging state $\mst$, namely $\mst(\pid{q},\pid{p})$, and the transition takes the communication to a state with a pending receive $\genrtcom$. In \textsc{recv-val}, $\pid{q}$ consumes the value at the head of that queue, stores it in its local variable $x$, and continues with $C$.
Communications over selection labels use the same mechanism, as shown in \textsc{send-sel} and \textsc{recv-sel}.

For modelling choreographic nondeterminism, rules \textsc{choice-l} and \textsc{choice-r} resolve a choreographic
choice by taking a transition in one of its branches. For instance,
if $C_1$ takes a $\mu$-transition to $C_1'$ and
$\pid{p}\in\pn(\mu)$, then \textsc{choice-l} records the decision in
$\tlleft{p}{\mu}$ and continues with $C_1'\fatsemi C$.
For a communication from $\pid{p}$ to $\pid{q}$, the send label has
process set $\{\pid{p}\}$ and the receive label has process set
$\{\pid{q}\}$. A choice at $\pid{q}$ can therefore be resolved by
receiving the message.
\begin{figure}[b]
\vspace{-1.0em}
\begin{mathpar}
    \mathshade{
    \inferrule*[right={\scriptsize send-val}] { \cstore(\pid{p}) \vdash e \downarrow v \\ \mst(\pid q, \pid p) = \vec{\msg} } 
    {\langle \gencom \seq C, \cstore, \mst, \mathscr{C} \rangle \xrightarrow[]{\tlcoms{p}{v}{q}} \langle \genrtcom \seq C, \cstore, \mst[(\pid{q},\pid{p}) \mapsto \vec{m} \cons v], \mathscr{C} \rangle}
    }

    \mathshade{
    \inferrule*[right={\scriptsize recv-val}] { \mst(\pid q, \pid p) = v \cons \vec{\msg} } 
    {\langle \genrtcom \seq C, \cstore, \mst, \mathscr{C} \rangle \xrightarrow[]{\tlcomr{p}{v}{q}} \langle C, \cstore[\pid{q}.x \mapsto v], \mst[(\pid{q},\pid{p}) \mapsto \vec{m}], \mathscr{C} \rangle}
    }

    \mathshade{
    \inferrule*[right={\scriptsize send-sel}] { \mst(\pid q, \pid p) = \vec{\msg} } 
    {\langle \gensel \seq C, \cstore, \mst, \mathscr{C} \rangle \xrightarrow[]{\tlsels{p}{q}{\albl}} \langle \genrtsel \seq C, \cstore, \mst[(\pid{q},\pid{p}) \mapsto \vec{m} \cons \albl], \mathscr{C} \rangle}
    }

    \mathshade{
    \inferrule*[right={\scriptsize recv-sel}] { \mst(\pid q, \pid p) = \albl \cons \vec{\msg} } 
    {\langle \genrtsel \seq C, \cstore, \mst, \mathscr{C} \rangle \xrightarrow[]{\tlselr{p}{q}{\albl}} \langle C, \cstore, \mst[(\pid{q},\pid{p}) \mapsto \vec{m}], \mathscr{C} \rangle}
    }
    
    \mathshade{
    \inferrule*[right={\scriptsize choice-l}]{\langle C_1, \cstore, \mst, \mathscr{C}\rangle \xrightarrow[]{\mu} \langle C_1', \cstore', \mst', \mathscr{C}\rangle \\ \pid{p}\in \pn(\mu)}{\langle C_1 \choice_\pid{p} C_2\seq C, \cstore, \mst, \mathscr{C}\rangle \xrightarrow[]{\tlleft{p}{\mu}} \langle C_1' \fatsemi C, \cstore', \mst', \mathscr{C} \rangle}
    }
    
    \mathshade{
    \inferrule*[right={\scriptsize choice-r}]{\langle C_2, \cstore, \mst, \mathscr{C}\rangle \xrightarrow[]{\mu} \langle C_2', \cstore', \mst', \mathscr{C}\rangle \\ \pid{p}\in \pn(\mu)} { \langle C_1 \choice_\pid{p} C_2\seq C, \cstore, \mst, \mathscr{C}\rangle \xrightarrow[]{\tlright{p}{\mu}} \langle C_2' \fatsemi C, \cstore', \mst', \mathscr{C} \rangle }
    }
\end{mathpar}
\caption{Semantics of Choreographies in \async, Rules Concerning Asynchrony and Nondeterminism}
\label{chor:semantics:lts:fig-1}
\vspace{-1.0em}
\end{figure}

\begin{figure}
\vspace{-1.0em}
\begin{mathpar}
    \inferrule*[right={\scriptsize assign}] { \cstore(\pid{p}) \vdash e \downarrow v }
    {\langle \pid{p}.x \metaDef e \seq C, \cstore, \mst, \mathscr{C} \rangle \xrightarrow[]{\tlassign{\pid{p}}} \langle C, \cstore[\pid{p}.x \mapsto v], \mst, \mathscr{C} \rangle}

    \inferrule*[right={\scriptsize cond-then}] { \cstore(\pid{p}) \vdash e \downarrow \mathit{true}} 
    {\langle \condif \pid{p}.e\condthen C_1\condelse C_2 \seq C, \cstore, \mst, \mathscr{C} \rangle \xrightarrow[]{\tlthen{p}} \langle C_1 \fatsemi C, \cstore, \mst, \mathscr{C} \rangle }

    \inferrule*[right={\scriptsize cond-else}] { \cstore(\pid{p}) \vdash e \downarrow v \\ v \neq \mathit{true}} 
    {\langle \condif \pid{p}.e\condthen C_1\condelse C_2 \seq C, \cstore, \mst, \mathscr{C} \rangle \xrightarrow[]{\tlelse{p}} \langle C_2 \fatsemi C, \cstore, \mst, \mathscr{C} \rangle }

    \inferrule*[right={\scriptsize call-first}] { X(\pids{q}) = C \in \mathscr{C} \\ \pids{p} = \pid{p}_1, \ldots, \pid{p}_n \\ i \in [1,n] }
    {\langle X(\pids{p});C', \cstore, \mst, \mathscr{C} \rangle \xrightarrow[]{\tlcall{\pid{p}_i}} \\\langle \pid{p}_1 : X(\pids{p}).C'; \ldots; \pid{p}_{i-1} : X(\pids{p}).C'; \\ \pid{p}_{i+1} : X(\pids{p}).C'; \ldots; \pid{p}_n : X(\pids{p}).C'; (C[\pids{p}/\pids{q}] \fatsemi C'), \cstore, \mst, \mathscr{C} \rangle}

    \inferrule*[right={\scriptsize call-enter}] {  }
    {\langle \pid{q} : X(\pids{p}).C'; C, \cstore, \mst, \mathscr{C} \rangle \xrightarrow[]{\tlcall{\pid{q}}} \langle C, \cstore, \mst, \mathscr{C} \rangle}
    
    \inferrule*[right={\scriptsize skip-l}]{\pid{p} \notin \pn(C_1) \cup \pn(C_2)}{ \langle C_1 \choice_\pid{p} C_2\seq C, \cstore, \mst, \mathscr{C}\rangle \xrightarrow[]{\tlskip{p}} \langle C_1 \fatsemi C, \cstore, \mst, \mathscr{C} \rangle }
    
    \inferrule*[right={\scriptsize skip-r}]{\pid{p} \notin \pn(C_1) \cup \pn(C_2)}{ \langle C_1 \choice_\pid{p} C_2\seq C, \cstore, \mst, \mathscr{C}\rangle \xrightarrow[]{\tlskip{p}} \langle C_2 \fatsemi C, \cstore, \mst, \mathscr{C} \rangle }

    \inferrule*[right={\scriptsize delay}] {\langle C, \cstore, \mst, \mathscr{C}\rangle \xrightarrow[]{\mu} \langle C', \cstore', \mst', \mathscr{C}\rangle \\ \pn(I)\disj\pn(\mu)}
    {\langle I\seq C, \cstore, \mst, \mathscr{C}\rangle \xrightarrow[]{\mu} \langle I\seq C', \cstore', \mst', \mathscr{C}\rangle}

    \inferrule*[right={\scriptsize delay-cond}] {\langle C_1, \cstore, \mst, \mathscr{C}\rangle \xrightarrow[]{\mu} \langle C_1', \cstore', \mst', \mathscr{C}\rangle \\ \langle C_2, \cstore, \mst, \mathscr{C}\rangle \xrightarrow[]{\mu} \langle C_2', \cstore', \mst', \mathscr{C}\rangle \\ \pid{p}\notin \pn(\mu)}
    {\langle \condif \pid{p}.e\condthen C_1\condelse C_2 \seq C, \cstore, \mst, \mathscr{C} \rangle \xrightarrow[]{\mu} \langle \condif \pid{p}.e\condthen C_1'\condelse C_2' \seq C, \cstore', \mst', \mathscr{C} \rangle}
    
    \inferrule*[right={\scriptsize delay-choice}]{\langle C_1, \cstore, \mst, \mathscr{C}\rangle \xrightarrow[]{\mu} \langle C_1', \cstore', \mst', \mathscr{C}\rangle \\ \langle C_2, \cstore, \mst, \mathscr{C}\rangle \xrightarrow[]{\mu} \langle C_2', \cstore', \mst', \mathscr{C}\rangle \\ \pid{p} \notin \pn(\mu)}{ \langle C_1 \choice_\pid{p} C_2\seq C, \cstore, \mst, \mathscr{C}\rangle \xrightarrow[]{\mu} \langle C_1' \choice_\pid{p} C_2'\seq C, \cstore', \mst', \mathscr{C} \rangle }
\end{mathpar}
\caption{Semantics of Choreographies in \async, Remaining Rules}
\label{chor:semantics:lts:fig-2}
\vspace{-1.0em}
\end{figure}

\Cref{chor:semantics:lts:fig-2} gives the remaining rules. An assignment
updates the local store, and a conditional evaluates its guard expression
at the guard process to choose a branch. In \cref{ex:CinvC}, for example,
the shop $\pid{p}$ evaluates $\fname{pay()}$ to decide whether to take
the payment branch.
Rules \textsc{call-first} and \textsc{call-enter} model general recursion. Rule \textsc{call-first} unfolds the procedure body with the actual process arguments and records the other participants as pending. Here, We write $C[\pids{p}/\pids{q}]$ for simultaneous substitution of process names in a choreography (cf. definitions~\ref{chor:semantics:pid-pids-substitution} and \ref{chor:semantics:chor-substitution}). Each of those participants enters by removing its own marker via \textsc{call-enter}. The garbage collection rules \textsc{skip-l} and \textsc{skip-r} are important for ensuring deadlock freedom. They permit either branch when the choice-resolving process is absent from both branches.

We have three rules for modelling out-of-order executions. Rule
\textsc{delay} allows the continuation $C$ to perform a $\mu$-transition
when the active processes in $\mu$ are not mentioned in $\pn(I)$. For instance, in \cref{ex:reusable-lock}, after $\pid{s}$ sends $\lbl{ok}$ to $\pid{p_1}$, \textsc{delay} allows $\pid{s}$ to take the snapshot while the reply is queued.
Rules \textsc{delay-cond} and \textsc{delay-choice} allow both branches to take a transition with the same label $\mu$.
The guard process or choice-resolving process must be absent from $\pn(\mu)$, so the conditional or choice remains pending. Using the rules \textsc{delay} and \textsc{delay-choice}, together with \textsc{choice-l} and \textsc{choice-r}, we can express interesting transitions for the lock acquisition in \cref{ex:reusable-lock}. 

Consider one unfolded iteration in which $\pid{p_1}$ sends its request
before $\pid{p_2}$, but $\pid{p_2}$'s request is received by $\pid{s}$
first, leading to $\pid{p_2}$ winning the competition.
Let $C_1$ be the left branch and $C_2$ the right branch. Both request
queues are initially empty in $\mst_0$.
By \textsc{delay}, each send can pass instructions involving only the
other client and $\pid{s}$. Each send thus produces the same label and
queue update in both branches. Since $\pid{s}$ is absent from the
process set of either send label, \textsc{delay-choice} advances both
branches and leaves the choice unresolved. The two sends update separate queues. The sends and the subsequent receive produce the following messaging states:
$\mst_1=\mst_0[(\pid{s},\pid{p_1})\mapsto[\lbl{req}]],\;
\mst_2=\mst_1[(\pid{s},\pid{p_2})\mapsto[\lbl{req}]],\;
\mst_3=\mst_2[(\pid{s},\pid{p_2})\mapsto[]]=\mst_1 .$

The corresponding choreographic transitions for this round of competition are shown below:
\begin{spreadlines}{0pt}
\begin{align*}
&\langle (C_1\choicep{s}C_2)\seq X(\pid{s},\pid{p_1},\pid{p_2}),
  \cstore,\mst_0,\mathscr{C}\rangle
 \xrightarrow{\tlsels{p_1}{s}{req}}
 \langle (C_1'\choicep{s}C_2')\seq X(\pid{s},\pid{p_1},\pid{p_2}),
  \cstore,\mst_1,\mathscr{C}\rangle
 \xrightarrow{\tlsels{p_2}{s}{req}}&&\\
&\quad\langle (C_1''\choicep{s}C_2'')\seq X(\pid{s},\pid{p_1},\pid{p_2}),
  \cstore,\mst_2,\mathscr{C}\rangle
 \xrightarrow{\tlright{s}{\tlselr{p_2}{s}{req}}}
 \langle C_{2\mathrm{win}}\fatsemi X(\pid{s},\pid{p_1},\pid{p_2}),
  \cstore,\mst_3,\mathscr{C}\rangle .&&
\end{align*}
\end{spreadlines}
The first send yields $C_1'$ and $C_2'$, and the second yields
$C_1''$ and $C_2''$. When $\pid{s}$ receives $\pid{p_2}$'s request,
\textsc{choice-r} selects the right branch.
The request from $\pid{p_1}$ remains queued, with its pending receive following $W(\pid{p_2},\pid{s})$ in $C_{2\mathrm{win}}$. The definitions below show the resulting choreographies in each transition:
\begin{spreadlines}{0pt}
\renewcommand{\seq}{;}
\thinmuskip=2mu
\medmuskip=1mu
\thickmuskip=2mu
\begin{flalign*}
&C_1' \defeq \rtsel{p_1}{s}{req}\seq \csel{s}{p_1}{ok}\seq \assign{s}{\vname{backup}}{\fname{snapshot}(\vname{inventory})}\seq W(\pid{p_1},\pid{s})\seq \csel{p_2}{s}{req}\seq \csel{s}{p_2}{ko},&&\displaybreak[0]\\[0.5ex]
&C_2' \defeq \csel{p_2}{s}{req}\seq \csel{s}{p_2}{ok}\seq \assign{s}{\vname{backup}}{\fname{snapshot}(\vname{inventory})}\seq W(\pid{p_2},\pid{s})\seq \rtsel{p_1}{s}{req}\seq \csel{s}{p_1}{ko},&&\\[0.5ex]
&C_1'' \defeq \rtsel{p_1}{s}{req}\seq \csel{s}{p_1}{ok}\seq \assign{s}{\vname{backup}}{\fname{snapshot}(\vname{inventory})}\seq W(\pid{p_1},\pid{s})\seq \rtsel{p_2}{s}{req}\seq \csel{s}{p_2}{ko},&&\\[0.5ex]
&C_2'' \defeq \rtsel{p_2}{s}{req}\seq \csel{s}{p_2}{ok}\seq \assign{s}{\vname{backup}}{\fname{snapshot}(\vname{inventory})}\seq W(\pid{p_2},\pid{s})\seq \rtsel{p_1}{s}{req}\seq \csel{s}{p_1}{ko},&&\\[0.5ex]
&C_{2\mathrm{win}} \defeq \csel{s}{p_2}{ok}\seq \assign{s}{\vname{backup}}{\fname{snapshot}(\vname{inventory})}\seq W(\pid{p_2},\pid{s})\seq \rtsel{p_1}{s}{req}\seq \csel{s}{p_1}{ko}.&&
\end{flalign*}
\end{spreadlines}

Sequential composition creates a boundary of executions, Transitions never cross a sequential composition operator.
This property is fundamentally important for distinguishing the \textsc{delay}, \textsc{delay-cond}, and \textsc{delay-choice} cases for establishing the behavioural equivalence between out-of-order executions in choreographies and parallel executions in networks \cref{sec:epp:correctness}.
\begin{theorem}[Execution Does Not Cross Sequential Composition]
\label{chor:semantics:seqcomp-border}
For any choreographic configuration $\langle C_1 \fatsemi C_2, \cstore, \mst, \mathscr{C} \rangle$, if it can perform a $\mu$-transition and \emph{$\pn(\mu) \subseteq \pn(C_1)$}, then there exists a choreography $C_1'$ such that $\langle C_1, \cstore, \mst, \mathscr{C} \rangle \xrightarrow[]{\mu} \langle C_1', \cstore', \mst', \mathscr{C} \rangle$ and the resulting choreographic configuration after the $\mu$-transition performed by $\langle C_1 \fatsemi C_2, \cstore, \mst, \mathscr{C} \rangle$ is $\langle C_1' \fatsemi C_2, \cstore', \mst', \mathscr{C} \rangle$.
\end{theorem}
\begin{proof}
    By structural induction on the derivation of the transition.
\end{proof}

\subsection{Messaging State Consistency and Well-Formedness}
\label{chor:wf}
We first introduce a new concept, \emph{messaging state consistency}, which specifies that all values and labels in message queues should be consumed by runtime terms. We write $\mst \cst C$ to say $\mst$ is consistent with $C$. Formally, we define messaging state consistency in \cref{chor:queue-consistency:fig}.
\begin{figure}[b]
\begin{mathpar}
\inferrule*[right={\scriptsize con-end}]{ \forall\, \pid{p}\, \pid{q},\, \mst (\pid{q}, \pid{p}) = []}{\mst \cst \cnil}

\inferrule*[right={\scriptsize con-local}]{\mst \cst C}{\mst \cst \pid{p}.x \metaDef e \seq C}

\inferrule*[right={\scriptsize con-com}]{\mst (\pid{q}, \pid{p}) = [] \\ \mst \cst C}{\mst \cst \gencom \seq C}

\inferrule*[right={\scriptsize con-rt-com}]{\mst (\pid{q}, \pid{p}) = v \cons \vec{\msg} \\ \mst[(\pid q, \pid p) \mapsto \vec{m}] \cst C}{\mst \cst \genrtcom \seq C}

\inferrule*[right={\scriptsize con-sel}]{\mst (\pid{q}, \pid{p}) = [] \\ \mst \cst C}{\mst \cst \gensel \seq C}

\inferrule*[right={\scriptsize con-rt-sel}]{\mst (\pid{q}, \pid{p}) = \albl \cons \vec{\msg} \\ \mst[(\pid q, \pid p) \mapsto \vec{m}] \cst C}{\mst \cst \genrtsel \seq C}

\inferrule*[right={\scriptsize con-cond}]{\mst \cst (C_1 \fatsemi C) \\ \mst \cst (C_2 \fatsemi C)}{\mst \cst \condif \pid{p}.e\condthen C_1\condelse C_2 \seq C}

\inferrule*[right={\scriptsize con-choice}]{\mst \cst (C_1 \fatsemi C) \\ \mst \cst (C_2 \fatsemi C)}{\mst \cst  C_1 \choice_\pid{p} C_2 \seq C}

\inferrule*[right={\scriptsize con-call}]{\forall\, \pid{p}\, \pid{q} \in \pids{p},\, \mst (\pid{q}, \pid{p}) = []\\ \mst \cst C}{\mst \cst X(\pids{p}) \seq C}

\inferrule*[right={\scriptsize con-rt-call}]{\mst \cst C}{\mst \cst \pid{q} : X(\pids{p}).C' \seq C}
\end{mathpar}
\caption{Messaging State Consistency in \async}
\label{chor:queue-consistency:fig}
\end{figure}

We also define the standard \emph{well-formedness} checks for choreographies and configurations to exclude erroneous situations such as a process communicating with itself or incorrect usage of procedure parameters.
The well-formedness judgement $\langle \mathscr{C},\{\pids{r}\}, C \rangle \checkmark$ states that a choreography $C$ is well-formed w.r.t. a set of choreographic procedure definitions $\mathscr{C}$ and a set of participating processes $\{\pids{r}\}$ (see \cref{app:chor:wf}).
In particular, each communication or selection, including its runtime form, must name two distinct processes from $\{\pids{r}\}$.
For each procedure call, including its runtime form, the arguments must be distinct processes from $\{\pids{r}\}$.
The called procedure must be defined, and the number of arguments must match the number of parameters in its definition.

We give the formal definition of the \emph{well-formed choreographic procedure definitions $\mathscr{C}$} as a predicate over $\mathscr{C}$ and the \emph{well-formed choreographic configurations} in definitions~\ref{chor:wf:procs:def} and \ref{chor:wf:config:def}.
\begin{definition}[Well-Formed Choreographic Procedure Definitions]\label{chor:wf:procs:def}
A set of choreographic procedure definitions $\mathscr{C}$ is well-formed if for each $X(\pids{p}) = C$ in $\mathscr{C}$,
all processes in $\pids{p}$ are distinct, $C$ does not contain any runtime terms, and $\langle\mst, \mathscr{C}, \{\pids{p}\}, C \rangle \checkmark$.
\end{definition}

\begin{definition}[Well-Formed Choreographic Configurations]\label{chor:wf:config:def}
    A configuration $\langle C, \cstore, \mst, \mathscr{C} \rangle$ is well-formed if both $\mathscr{C}$ and $C$ are well-formed, and $\mst \cst C$.
\end{definition}

\Cref{chor:wf:preservation:thm} states that well-formedness is preserved by transitions, which is crucial for proving the behavioural equivalence between choreographies and their projections in section~\ref{sec:epp:correctness}.
\begin{theorem}[Preservation of Well-Formedness]\label{chor:wf:preservation:thm}
    Given a well-formed choreographic configuration $\langle C, \cstore, \mst, \mathscr{C} \rangle$, if $\langle C, \cstore, \mst, \mathscr{C} \rangle \xrightarrow[]{\mu} \langle C', \cstore', \mst', \mathscr{C} \rangle$, then $\langle C', \cstore', \mst', \mathscr{C} \rangle$ is also well-formed.
\end{theorem}
\begin{proof}
    By structural induction on the derivation of the transition.
\end{proof}

%% file: sections/process-network.tex
\section{Processes and Networks}
\label{sec:proc-net}
\emph{Networks} are collections of \emph{endpoint programs}, and
endpoint programs are expressed in the \emph{process language}. We adopt the syntax of the process language and networks in \mech, and design new network semantics to model nondeterminism and asynchrony.
\subsection{Syntax}
\label{sec:proc-net:syntax}
\Cref{proc:syntax:fig} is the formal syntax of the process language. 
We mutually define terms $P$ and instructions $I$, and overload $I$ for process instructions.
\begin{figure}
\vspace{-1.0em}
\begin{spreadlines}{0pt}
\begin{alignat*}{3}
    \text{Process Procedure Definitions ($\mathfrak{P}$)}\ &&
    \mathscr{P} \metaDeff&\ \{X_i(\pids{p_i}) = P_i\}_{i \in \idx{I}} \\
    \text{Process Term (\setof{Process})}\ &&
    P, Q \metaDeff&\ I \seq P \cmid \precvl p \cmid \pnil\\
    \text{Process Instruction (\setof{PInstr})}\ &&
    I \metaDeff&\ x \metaDef e \cmid \psend{p}{e} \cmid \precv{p}{x} \cmid \psel{p} \cmid \condif e \condthen P\condelse Q \\ && \cmid &\ P + Q \cmid X(\pids{p})
\end{alignat*}
\end{spreadlines}
\caption{Syntax of the Process Language in \async}
\label{proc:syntax:fig}
\vspace{-1.0em}
\end{figure}
A \emph{network} is a \emph{function} mapping process names to process terms, where the zero element of the codomain \setof{Process} is the terminated process $\pnil$.
\[ 
\mathrm{Network\;(\setof{Net})}\quad N : \setof{PName} \to \setof{Process}
\]
Since a network is a function, we establish the equality of two networks using \emph{extensional equality}.
Formally, for two networks $N$ and $M$, $N = M$ if for all processes $\pid{p}$,  $N(\pid{p}) = M(\pid{p})$.
The support $\support(N)$ consists of the processes \emph{running} in $N$, namely those for which $N(\pid{p}) \ne \pnil$ (cf. \cref{sec:finfun}).
We write $N \disjoint M$ when their supports are disjoint.
As in \mech, the terminated network $\nnil$ maps every process to $\pnil$, and the atomic network $\pid{p}[P]$ maps $\pid{p}$ to a non-terminated process term $P$ and every other name to $\pnil$. Parallel composition $N \parcomp M$ is left-biased and right-associative:
\begin{spreadlines}{0pt}
\begin{align*}
\begin{aligned}[c]
    \nnil (\pid{p}) \triangleq \pnil
\end{aligned}
\quad\quad
\begin{aligned}[c]
    \pid{p}[P] (\pid {q})&\triangleq 
    \begin{cases}
    \ P &\text{if $\pid{p} = \pid{q}$}\\[-0.5ex]
    \ \pnil &\text{otherwise}
    \end{cases}
\end{aligned}
\quad
\begin{aligned}[c]
    (N \parcomp M)(\pid {p})&\triangleq 
    \begin{cases}
    \ N(\pid{p}) &\text{if $\pid{p} \in \support(N)$}\\[-0.5ex]
    \ M(\pid{p}) &\text{otherwise}
    \end{cases}
\end{aligned}
\end{align*}
\end{spreadlines}
Networks form a \emph{monoid} under $\parcomp$ with identity $\nnil$ (see \cref{app:proc:parcomp:prop}).
\begin{figure}[b]
\vspace{-1.0em}
\begin{align*}
\begin{aligned}[c]
\pnil \fatsemi P &\triangleq P
\end{aligned}
\quad\quad\quad
\begin{aligned}[c]
(I\seq P') \fatsemi P &\triangleq I \seq (P' \fatsemi P)
\end{aligned}
\quad\quad\quad
\begin{aligned}[c]
\precvl p \fatsemi P &\triangleq \pid{p} \,\&\,\{\albl_i : P_i \fatsemi P\}_{i \in \idx{I}}
\end{aligned}
\end{align*}
\vspace{-1.7em}
\begin{align*}
(N \fatsemi M)\, (\pid{p}) \triangleq (N\, (\pid{p}))\fatsemi (M \, (\pid{p}))
\end{align*}
\caption{Sequential Composition of Process Terms and Networks}
\label{proc:semantics:seqcomp}
\vspace{-1.0em}
\end{figure}

\subsection{Semantics}
\label{sec:proc-net:semantics}
\begin{figure}[b]
\vspace{-1.0em}
\begin{mathpar}
    \inferrule*[right={\scriptsize assign}] { \cstore(\pid{p}) \vdash e \downarrow v }
    {\langle \pid{p}[x \metaDef e \seq P], \cstore, \mst, \mathscr{P} \rangle \xrightarrow[]{\tlassign{\pid{p}}} \langle \pid{p}[P], \cstore[\pid{p}.x \mapsto v], \mst, \mathscr{P} \rangle}

    \mathshade{
    \inferrule*[right={\scriptsize send-val}] { \cstore(\pid{p}) \vdash e \downarrow v \\ \mst (\pid q, \pid p) = \vec{m} } 
    {\langle \pid{p}[\psend{q}{e} \seq P], \cstore, \mst, \mathscr{P} \rangle \xrightarrow[]{\tlcoms{p}{v}{q}} \langle \pid{p}[P], \cstore, \mst [(\pid q, \pid p) \mapsto \vec{m} \cons v], \mathscr{P} \rangle}
    }

    \mathshade{
    \inferrule*[right={\scriptsize recv-val}] { \mst (\pid q, \pid p) = v \cons \vec{m} } 
    {\langle \pid{q}[\precv{p}{x} \seq Q], \cstore, \mst, \mathscr{P} \rangle \xrightarrow[]{\tlcomr{p}{v}{q}} \langle \pid{q}[Q], \cstore[\pid{q}.x \mapsto v], \mst[(\pid q, \pid p) \mapsto \vec{m}], \mathscr{P} \rangle}
    }

    \mathshade{
    \inferrule*[right={\scriptsize send-sel}] { \mst (\pid q, \pid p) = \vec{m} } 
    {\langle \pid{p}[\psel{q} \seq P], \cstore, \mst, \mathscr{P} \rangle \xrightarrow[]{\tlsels{p}{q}{$\albl$}} \langle \pid{p}[P], \cstore, \mst [(\pid q, \pid p) \mapsto \vec{m} \cons \albl], \mathscr{P} \rangle}
    }

    \mathshade{
    \inferrule*[right={\scriptsize recv-sel}] { j \in \mathcal{I} \\ \mst (\pid q, \pid p) = \albl_j \cons \vec{m} } 
    {\langle \pid{q}[\precvl p ], \cstore, \mst, \mathscr{P} \rangle \xrightarrow[]{\tlselr{p}{q}{$\albl_j$}} \langle \pid{q}[P_j], \cstore, \mst [(\pid q, \pid p) \mapsto \vec{m}], \mathscr{P} \rangle}
    }

    \inferrule*[right={\scriptsize cond-then}] { \cstore(\pid{p}) \vdash e \downarrow \mathit{true}} 
    {\langle \pid{p}[\condif e \condthen P_1 \condelse P_2 \seq Q], \cstore, \mst, \mathscr{P} \rangle \xrightarrow[]{\tlthen{p}} \langle \pid{p} [P_1 \fatsemi Q], \cstore, \mst, \mathscr{P} \rangle }

    \inferrule*[right={\scriptsize cond-else}] { \cstore(\pid{p}) \vdash e \downarrow v \\ v \neq \mathit{true}} 
    {\langle \pid{p}[\condif e \condthen P_1 \condelse P_2 \seq Q], \cstore, \mst, \mathscr{P} \rangle \xrightarrow[]{\tlelse{p}} \langle \pid{p} [P_2 \fatsemi Q], \cstore, \mst, \mathscr{P} \rangle }
    
    \mathshade{
    \inferrule*[right={\scriptsize choice-l}]{\langle \pid{p}[P], \cstore, \mst, \mathscr{P}\rangle \xrightarrow[]{\mu} \langle \pid{p}[P'], \cstore', \mst', \mathscr{P}\rangle \\ \pid{p} \in \pn(\mu)} {\langle \pid{p}[P \choice Q \seq R], \cstore, \mst, \mathscr{P}\rangle \xrightarrow[]{\tlleft{p}{\mu}} \langle\pid{p}[P' \fatsemi R], \cstore', \mst', \mathscr{P} \rangle}
    }
    
    \mathshade{
    \inferrule*[right={\scriptsize choice-r}]{\langle \pid{p}[Q], \cstore, \mst, \mathscr{P}\rangle \xrightarrow[]{\mu} \langle \pid{p}[Q'], \cstore', \mst', \mathscr{P}\rangle \\ \pid{p} \in \pn(\mu)} {\langle \pid{p}[P \choice Q \seq R], \cstore, \mst, \mathscr{P}\rangle \xrightarrow[]{\tlright{p}{\mu}}\langle\pid{p}[Q' \fatsemi R], \cstore', \mst', \mathscr{P} \rangle}
    }

    \inferrule*[right={\scriptsize choice-skip}] { } 
    {\langle \pid{p}[\cnil + \cnil \seq P], \cstore, \mst, \mathscr{P} \rangle \xrightarrow[]{\tlskip{p}} \langle \pid{p} [P], \cstore, \mst, \mathscr{P} \rangle }

    \inferrule*[right={\scriptsize call}] {X(\pids{r}) = Q \in \mathscr{P}} 
    {\langle \pid{p}[X(\pids{q}) \seq P], \cstore, \mst, \mathscr{P} \rangle \xrightarrow[]{\tlcall{p}} \langle \pid{p} [Q[\pids{q}/\pids{r}] \fatsemi P], \cstore, \mst, \mathscr{P} \rangle }

    \inferrule*[right={\scriptsize par}] {\langle N, \cstore,\mst, \mathscr{P} \rangle \xrightarrow[]{\mu} \langle N', \cstore', \mst', \mathscr{P} \rangle \\ N \disjoint M} 
    {\langle N \parcomp M, \cstore, \mst, \mathscr{P} \rangle \xrightarrow[]{\mu} \langle N' \parcomp M, \cstore', \mst', \mathscr{P} \rangle }
\end{mathpar}
\caption{Semantics of Networks in \async}
\label{proc:net:semantics:fig}
\vspace{-1.0em}
\end{figure}
We define the semantics of \emph{network configurations} $\langle N, \cstore, \mst, \mathscr{P} \rangle$ as an LTS. Transitions take the form of $\langle N, \cstore, \mst, \mathscr{P} \rangle \xrightarrow[]{\mu} \langle N', \cstore', \mst', \mathscr{P} \rangle$,
where $\mst$ is the messaging state in \cref{chor:semantics:messaging-state},
$\mu$ is a transition label of the same form as in the choreographic semantics (\cref{chor:semantics:transition-labels}).
This is important for establishing the behavioural equivalence between choreographies and their projected networks in \cref{sec:epp:correctness}.

\subsubsection{Monoidal Construction of Sequential Composition}
\label{sec:proc:semantics:seqcomp}
In \cref{proc:semantics:seqcomp} we formally define the sequential composition operation for process terms and networks, overloading the notation $(\fatsemi)$.
We define sequential composition for networks as the pointwise extension of sequential composition for process terms. The set of process terms and the set of networks each form a monoid under sequential composition. (cf. propositions~\ref{app:proc:seqcomp:prop} and \ref{app:net:seqcomp:prop}).

\subsubsection{LTS for Networks}
Unlike the synchronous network semantics, the asynchronous network semantics is local.
\Cref{proc:net:semantics:fig} gives the formal semantics for networks. The rule \textsc{assign} models local assignments, which update the executing process's local store.
Rules \textsc{send-val}, \textsc{recv-val}, \textsc{send-sel}, and \textsc{recv-sel} model asynchronous communication. 
In \textsc{send-val}, $\pid{p}$ evaluates $e$ in its local store, appends the resulting value $v$ to the FIFO queue $\mst(\pid{q},\pid{p})$, and continues with $P$ independently of $\pid{q}$. In \textsc{recv-val}, $\pid{q}$
removes a value $v$ from the head of that queue, stores it in its local variable $x$, and continues with $Q$.
In \textsc{send-sel}, $\pid{p}$ appends the selection label $\albl$ to the same queue and continues with $P$.
In \textsc{recv-sel}, when the head of the queue is a label $\albl_j$ offered by $\pid{q}$, the receiver removes it and continues with the corresponding branch $P_j$.

Rules \textsc{cond-then} and \textsc{cond-else} select $P_1$ if the local guard evaluates to $\mathit{true}$ and $P_2$ otherwise. Both rules then append the continuation $Q$.
Rules \textsc{choice-l} and \textsc{choice-r} resolve a local nondeterministic choice by taking a transition of one of its branches. If $P$ makes a $\mu$-transition to $P'$ and $\pid p \in \pn(\mu)$, \textsc{choice-l} resolves the choice to the left branch and continues with $P'\fatsemi R$.
Similarly, with a transition from $Q$ to $Q'$, \textsc{choice-r} resolves the choice to the right branch and
continues with $Q'\fatsemi R$. 
Both rules retain the store and queue updates from the branch transition. 
The projected reusable lock in \cref{ex:net-fcfs-rec} illustrates how a receive can resolve a choice in the same step. If both requests are queued, the server can resolve its choice by receiving either one.

The rule \textsc{choice-skip} is the network
counterpart of the choreographic garbage collection rules
\textsc{skip-l} and \textsc{skip-r}. It lets the process proceed
directly to the continuation when both branches are terminated.
Rule \textsc{call} models recursions. It substitutes the actual process names for the parameters in the procedure body and appends the caller's
continuation. We again write $P[\pids{q}/\pids{r}]$ for substituting process names in $P$ (cf. \cref{proc:semantics:proc-substitution}).
Rule \textsc{par} models parallel executions of disjoint networks. It allows one part of a network to make a transition while the other processes remain unaffected.

%% file: sections/nondeterminism-async.tex
\section{A Comparison of Synchrony and Asynchrony in the Presence of Nondeterminism}
\label{sec:nondeterminism-async}

In this section we compare how nondeterminism interacts with synchrony and asynchrony: we show that the former requires a cumbersome semantics and can express fewer behaviours, and we identify the lack of locality introduced by synchrony as the root cause.

\paragraph*{Synchrony complicates the semantics of nondeterministic choice} In \mech, a synchronous communication resolves the choices leading to the send and receive in a single synchronised transition, resulting in the non-local semantics in \cref{fig:mech-choice}. We show how nested choices is handled in \cref{ex:chor-nested-choice}.

\begin{example}[Nested Choreographic Choices and Their Projected Network]\label[example]{ex:chor-nested-choice}
Although the following choreographies $\cn{pq}$ and $\cn{qp}$ have different nesting structures:
\begin{spreadlines}{0pt}
\begin{align*}
\cn{pq} &\defeq (\ccom{p}{1}{q}{x} \choicep{q} \ccom{p}{1}{q}{y})
    \choicep{p} (\ccom{p}{2}{q}{x} \choicep{q} \ccom{p}{2}{q}{y})\\
\cn{qp} &\defeq (\ccom{p}{1}{q}{x} \choicep{p} \ccom{p}{2}{q}{x})
    \choicep{q} (\ccom{p}{1}{q}{y} \choicep{p} \ccom{p}{2}{q}{y}),
\end{align*}
\end{spreadlines}
they should exhibit the same behaviour, because they are projected into the same network $N$:
\[N \triangleq \atomicn{p}{\psend{q}{1} \choice \psend{q}{2}} \parcomp \atomicn{q}{\precv{p}{x} \choice \precv{p}{y}}.\]
Suppose $\pid{p}$ chooses left and $\pid{q}$ chooses right in one transition.

In \mech the derivation for $\cn{pq}$ resolves the choice at $\pid{q}$ before the choice at $\pid{p}$, while the derivation for $\cn{qp}$ resolves them in the opposite order. For EPP soundness, both derivations must have the same label, which also matches the corresponding transition of their common projected network. 
To give these derivations the same label, \mech pairs the communication label with a choice map from process names to lists of branch choices, as discussed in
\cref{chor:semantics:tl}. \mech's \textsc{choice-l} and \textsc{choice-r}
rules record the selected branch in the choice map for the process resolving the choice. Updates for distinct processes commute, so the two
derivations produce the same label.
\begin{spreadlines}{0pt}
\begin{align*}
\langle \cn{pq}, \cstore, \mathscr{C}\rangle
    &\xrightarrow{\addChoiceg{(\addChoiceg{\tlcom{p}{1}{q}}{q}{right})}{p}{left}}
    \langle \cnil, \cstore[\pid{q}.y \mapsto 1], \mathscr{C}\rangle\\[0.5ex]
\langle \cn{qp}, \cstore, \mathscr{C}\rangle
    &\xrightarrow{\addChoiceg{(\addChoiceg{\tlcom{p}{1}{q}}{p}{left})}{q}{right}}
    \langle \cnil, \cstore[\pid{q}.y \mapsto 1], \mathscr{C}\rangle .
\end{align*}
\end{spreadlines}
The network rules encode the synchronisation obligation of nondeterministic behaviours in communication by requiring a compatible joint step and recording the choices resolved in it:
\[
   \langle  N, \cstore, \mathscr{P} \rangle
    \xrightarrow[]{\addChoiceg{(\addChoiceg{\tlcom{p}{1}{q}}{p}{left})}{q}{right}}
    \langle \atomicn{p}{\pnil} \parcomp \atomicn{q}{\pnil},
    \cstore[\pid{q}.y \mapsto 1], \mathscr{P}\rangle.
\]

By contrast, the operational semantics of \async{} satisfies the \emph{locality property}: each transition has a single active process. We split communication into separate send and receive steps. Each process follows its own program and communicates through the queue in the messaging state. In the \async transitions, given $\mst_0(\pid{q},\pid{p}) = []$, $\pid{p}$ resolves its choice when it sends, and $\pid{q}$ resolves its choice when it receives. Each label therefore needs only one list of branch choices.
\begin{spreadlines}{0pt}
\begin{align*}
&\langle \cn{pq}, \cstore, \mst_0, \mathscr{C}\rangle
    \xrightarrow{\tlleft{p}{\tlcoms{p}{1}{q}}}
    \langle \rtcom{p}{q}{x} \choicep{q} \rtcom{p}{q}{y},
    \cstore, \mst_0[(\pid{q},\pid{p}) \mapsto [1]], \mathscr{C}\rangle\\[0.5ex]
&\langle \cn{qp}, \cstore, \mst_0, \mathscr{C}\rangle
    \xrightarrow{\tlleft{p}{\tlcoms{p}{1}{q}}}
    \langle \rtcom{p}{q}{x} \choicep{q} \rtcom{p}{q}{y},
    \cstore, \mst_0[(\pid{q},\pid{p}) \mapsto [1]], \mathscr{C}\rangle\\[0.5ex]
&\langle \rtcom{p}{q}{x} \choicep{q} \rtcom{p}{q}{y},
    \cstore, \mst_0[(\pid{q},\pid{p}) \mapsto [1]], \mathscr{C}\rangle
    \xrightarrow{\tlright{q}{\tlcomr{p}{1}{q}}}
    \langle \cnil, \cstore[\pid{q}.y \mapsto 1], \mst_0, \mathscr{C}\rangle .
\end{align*}
\end{spreadlines}
And the network transitions are:
\begin{spreadlines}{0pt}
\begin{align*}
\langle N, \cstore, \mst_0, \mathscr{P}\rangle
    &\xrightarrow{\tlleft{p}{\tlcoms{p}{1}{q}}}
    \langle \atomicn{p}{\pnil} \parcomp
    \atomicn{q}{\precv{p}{x} \choice \precv{p}{y}},
    \cstore, \mst_0[(\pid{q},\pid{p}) \mapsto [1]], \mathscr{P}\rangle\\
    &\xrightarrow{\tlright{q}{\tlcomr{p}{1}{q}}}
    \langle \atomicn{p}{\pnil} \parcomp \atomicn{q}{\pnil},
    \cstore[\pid{q}.y \mapsto 1], \mst_0, \mathscr{P}\rangle .
\end{align*}
\end{spreadlines}
\end{example}
\begin{figure}
\vspace{-1.0em}
\begin{mathpar}
    \inferrule*[right={\scriptsize choice-l (choreography)}]{
        \langle C_1, \cstore, \mathscr{C}\rangle \xrightarrow[]{\mu} \langle C_1', \cstore', \mathscr{C}\rangle \\
        \pid{p} \in \pn(\mu)
    }{
        \langle C_1 \choicep{p} C_2 \seq C, \cstore, \mathscr{C}\rangle
        \xrightarrow[]{\addChoice{p}{left}}
        \langle C_1' \fatsemi C, \cstore', \mathscr{C}\rangle
    }

    \inferrule*[right={\scriptsize choice-r (choreography)}]{
        \langle C_2, \cstore, \mathscr{C}\rangle \xrightarrow[]{\mu} \langle C_2', \cstore', \mathscr{C}\rangle \\
        \pid{p} \in \pn(\mu)
    }{
        \langle C_1 \choicep{p} C_2 \seq C, \cstore, \mathscr{C}\rangle
        \xrightarrow[]{\addChoice{p}{right}}
        \langle C_2' \fatsemi C, \cstore', \mathscr{C}\rangle
    }

    \inferrule*[right={\scriptsize choice-l (network)}]{
        \langle \pid{p}[P] \parcomp N, \cstore, \mathscr{P}\rangle
        \xrightarrow[]{\mu}
        \langle \pid{p}[P'] \parcomp N', \cstore', \mathscr{P}\rangle \\
        \pid{p} \in \pn(\mu) \\
        \pid{p} \notin \support(N) \\
        \pid{p} \notin \support(N') \\
        \support(N) \cup \{\pid{p}\} = \pn(\mu)
    }{
        \langle \pid{p}[P \choice Q \seq R] \parcomp N, \cstore, \mathscr{P}\rangle
        \xrightarrow[]{\addChoice{p}{left}}
        \langle \pid{p}[P' \fatsemi R] \parcomp N', \cstore', \mathscr{P}\rangle
    }

    \inferrule*[right={\scriptsize choice-r (network)}]{
        \langle \pid{p}[Q] \parcomp N, \cstore, \mathscr{P}\rangle
        \xrightarrow[]{\mu}
        \langle \pid{p}[Q'] \parcomp N', \cstore', \mathscr{P}\rangle \\
        \pid{p} \in \pn(\mu) \\
        \pid{p} \notin \support(N) \\
        \pid{p} \notin \support(N') \\
        \support(N) \cup \{\pid{p}\} = \pn(\mu)
    }{
        \langle \pid{p}[P \choice Q \seq R] \parcomp N, \cstore, \mathscr{P}\rangle
        \xrightarrow[]{\addChoice{p}{right}}
        \langle \pid{p}[Q' \fatsemi R] \parcomp N', \cstore', \mathscr{P}\rangle
    }
\end{mathpar}
\caption{Modelling Nondeterminism with Synchrony: Choice Rules for Choreographies and Networks in \mech}
\label{fig:mech-choice}
\vspace{-1.0em}
\end{figure}

\paragraph*{Synchrony expresses fewer behaviours.} With \async, in the execution
of the reusable lock (\cref{ex:reusable-lock}) shown in \cref{sec:chor},
$\pid{p_1}$ sends its request before $\pid{p_2}$. The requests enter separate
queues, and the server $\pid{s}$ in fact receives $\pid{p_2}$'s request
first, granting the lock to $\pid{p_2}$. In \mech, sending a
request and receiving it at $\pid{s}$ form a single synchronised
transition. If $\pid{p_1}$ sends first, $\pid{s}$ receives that
request in the same transition, resolving the choice in favour of
$\pid{p_1}$. Thus \mech ties the winner to the first
client to communicate its request. Using the same shorthand $C_1$ and $C_2$ for branches from \cref{sec:chor}, one unfolded iteration of the lock takes the following transitions:
\begingroup
\renewcommand{\seq}{;}
\thinmuskip=2mu
\medmuskip=1mu
\thickmuskip=2mu
\begin{align*}
&\langle (C_1 \choicep{s} C_2)\seq X(\pid{s},\pid{p_1},\pid{p_2}),
    \cstore,\mathscr{C}\rangle
    \xrightarrow{\addChoiceg{\csel{p_1}{s}{req}}{s}{left}}\\
&\langle \csel{s}{p_1}{ok}
    \seq \assign{s}{\vname{backup}}{\fname{snapshot}(\vname{inventory})}
    \seq W(\pid{p_1},\pid{s})
    \seq \csel{p_2}{s}{req}
    \seq \csel{s}{p_2}{ko}
    \seq X(\pid{s},\pid{p_1},\pid{p_2}),
    \cstore,\mathscr{C}\rangle .
\end{align*}
\endgroup

%% file: sections/epp.tex
\section{Endpoint Projection (EPP)}
\label{sec:epp}
EPP compiles choreographies into networks.
A central concern in the design of EPP is handling \emph{knowledge of choice} (KoC).
We address this using the select-and-merge approach.
Since our network language presents new semantics while leaving the syntax of the process terms and networks unchanged, we reuse the \emph{merge operation} ($\sqcup$) from \mech, as described in \cref{sec:background:merge}.

\subsection{Process Projection and EPP}
\label{epp:projection}
We define EPP using the \emph{process projection}, which produces the local program for a given process specified in a choreography. Process projection is a function of the type:
\[
\llbracket \cdot \rrbracket\_ : \textbf{Chor} \rightarrow \textbf{PName} \rightarrow \opt{\textbf{Process}}
\]
In \cref{epp:proc-projection:fig}, we show the algorithm of process projection.
\begin{figure}[b]
\vspace{-1.0em}
\begin{spreadlines}{1pt}
\begin{align*}
\begin{aligned}[c]
    \proj{\cnil}{r} &\triangleq \some{\pnil}
\end{aligned}
\quad\quad
\begin{aligned}[c]
    \proj{\pid{p}.x \metaDef e \seq C}{r} &\triangleq
    \begin{cases}
    \;\some{x \metaDef e \seq P} &\text{if $\pid{r} = \pid{p}$ and $\proj{C}{r} \eqr\some{P}$}\\[-0.5ex]
    \;\proj{C}{r} &\text{otherwise}
    \end{cases}
\end{aligned}
\end{align*}
\end{spreadlines}
\vspace{-1.0em}
\begin{spreadlines}{1pt}
\begin{align*}
\begin{aligned}[t]
    &\proj{\gencom \seq C}{r} \triangleq\\
    &\begin{cases}
    \;\some{\psend{q}{e} \seq P} &\text{if $\pid{r} = \pid{p}$ and $\proj{C}{r} \eqr\some{P}$}\\[-0.5ex]
    \;\some{\precv{p}{x} \seq P} &\text{if $\pid{r} = \pid{q}$ and $\proj{C}{r} \eqr\some{P}$}\\[-0.5ex]
    \;\proj{C}{r} &\text{otherwise}
    \end{cases}\\
    &\mathshade{
    \begin{aligned}[t]
    &\proj{\genrtcom \seq C}{r} \triangleq\\
    &\begin{cases}
    \;\some{\precv{p}{x} \seq P} &\text{if $\pid{r} = \pid{q}$ and $\proj{C}{r} \eqr\some{P}$}\\[-0.5ex]
    \;\proj{C}{r} &\text{otherwise}
    \end{cases}
    \end{aligned}
    }
\end{aligned}
\quad\quad\quad\quad\quad\quad\quad
\begin{aligned}[t]
    &\proj{\gensel \seq C}{r} \triangleq\\
    &\begin{cases}
    \;\some{\psel{q} \seq P} &\text{if $\pid{r} = \pid{p}$ and $\proj{C}{r} \eqr\some{P}$}\\[-0.5ex]
    \;\some{\precvlpr{p}{P}} &\text{if $\pid{r} = \pid{q}$ and $\proj{C}{r} \eqr\some{P}$}\\[-0.5ex]
    \;\proj{C}{r} &\text{otherwise}
    \end{cases}\\
    &\mathshade{
    \begin{aligned}[t]
    &\proj{\genrtsel \seq C}{r} \triangleq\\
    &\begin{cases}
    \;\some{\precvlpr{p}{P}} &\text{if $\pid{r} = \pid{q}$ and $\proj{C}{r} \eqr\some{P}$}\\[-0.5ex]
    \;\proj{C}{r} &\text{otherwise}
    \end{cases}
    \end{aligned}
    }
\end{aligned}
\end{align*}
\end{spreadlines}
\vspace{-1.0em}
\begin{spreadlines}{1pt}
\begin{align*}
    &\proj{\condif\pid{p}.e\condthen C_1\condelse C_2 \seq C}{r} \,\triangleq\\
    &\begin{cases}
    \;\some{\condif e\condthen P_1 \condelse P_2 \seq P} &\text{if $\pid{r} = \pid{p}$ and $\proj{C_1}{r} \eqr\some{P_1}$ and $\proj{C_2}{r} \eqr\some{P_2}$ and $\proj{C}{r} \eqr\some{P}$}\\[-0.5ex]
    \;\some{Q \fatsemi P} &\text{if $\pid{r} \neq \pid{p}$ and $\proj{C_1}{r} \sqcup \proj{C_2}{r} \eqr\some{Q}$ and $\proj{C}{r} \eqr\some{P}$}\\[-0.5ex]
    \;\none &\text{otherwise}
    \end{cases}\\
    &\proj{C_1 \choice_\pid{p} C_2 \seq C}{r} \,\triangleq
    \begin{cases}
    \;\some{P_1 + P_2 \seq P} &\text{if $\pid{r} = \pid{p}$ and $\proj{C_1}{r} \eqr\some{P_1}$ and $\proj{C_2}{r} \eqr\some{P_2}$ and $\proj{C}{r} \eqr\some{P}$}\\[-0.5ex]
    \;\some{Q \fatsemi P} &\text{if $\pid{r} \neq \pid{p}$ and $\proj{C_1}{r} \sqcup \proj{C_2}{r} \eqr\some{Q}$ and $\proj{C}{r} \eqr\some{P}$}\\[-0.5ex]
    \;\none &\text{otherwise}
    \end{cases}\\
    &\proj{X(\pids{p}) \seq C}{r} \,\triangleq
    \begin{cases}
    \;\some{X_i(\pids{p}\backslash \pid{r}) \seq P} &\text{if $\pid{r}=\pid{p}_i$ for $\pids{p} = \pid{p}_1,...,\pid{p}_n$ and $1\leq i\leq n$ and $\proj{C}{r} \eqr\some{P}$}\\[-0.5ex]
    \;\proj{C}{r} &\text{otherwise}
    \end{cases}\\
    &\proj{\pid{q} : X(\pids{p}). C' \seq C}{r} \,\triangleq
    \begin{cases}
    \;\some{X_i(\pids{p}\backslash \pid{r}) \seq P'} \!&\text{if $\pid{r} = \pid{q} =\pid{p}_i$ for $\pids{p} = \pid{p}_1,...,\pid{p}_n$, $1\leq i\leq n$, and $\proj{C'}{r} \eqr\some{P'}$}\\[-0.5ex]
    \;\llbracket C \rrbracket_\pid{r} \!&\text{otherwise}
    \end{cases}
\end{align*}
\end{spreadlines}
\caption{Process Projection of Choreographies in \async}
\label{epp:proc-projection:fig}
\vspace{-1.0em}
\end{figure}
A terminated choreography projects to a terminated process. We project a sequence $I \seq C$ by first projecting the choreographic instruction $I$. When $I$ is a local assignment, we project it to an assignment at the same process. A communication projects to a send and a receive, and a selection projects to a selection instruction and a branching term. For the runtime instructions $\genrtcom$ and $\genrtsel$, $\pid{p}$ has already sent the value or label. We therefore project only the pending action at $\pid{q}$, a receive into $x$ or a branching term, respectively. In each case, the projection of $C$ provides the continuation. At processes with no action for $I$, we project only $C$.

A conditional projects to a local conditional with the same guard expression at the guard process. A choreographic choice projects to a nondeterministic choice at the choice-resolving process. At each other process, we merge the projections of the two branches. We then append the projection of $C$. If a required projection or merge yields $\none$, the result is $\none$.
To project a choreographic procedure call onto a participating process $\pid{p}$, we use its position $i$ in the argument list to obtain the local procedure $X_i$. The projected call takes the other participating processes as arguments. After the call, $\pid{p}$ continues with its part of $C$, indicated by $\proj{C}{p}$. In the runtime term $\pid{q} : X(\pids{p}).C' \seq C$, only $\pid{q}$ has yet to enter the call, so we project the local call there and use the projection of $C'$ as its continuation. Otherwise, we project only $C$.
\begin{figure}
\vspace{-1.0em}
\begin{subfigure}{1.0\textwidth}
\begin{align*}
    \epp{C} &\triangleq
    \begin{cases}
        \,\some{\pid{p}\mapsto P_{\pid{p}}}&\text{if $\proj{C}{p} \eqr \some{P_{\pid{p}}}$ for each $\pid{p}\in\pn(C)$ and $P_{\pid{p}}=\pnil$ for each $\pid{p}\notin\pn(C)$}\\[-0.5ex]
        \,\none &\text{if there exists $\pid{p} \in \pn(C)$ such that $\proj{C}{p} \eqr \none$}
    \end{cases}
\end{align*}
\vspace{-1.0em}
\caption{Endpoint Projection for Choreographies}
\label{epp:chor:def:fig}
\end{subfigure}
\begin{subfigure}{1.0\textwidth}
\begin{align*}
    \epp{\mathscr{C}} &\triangleq
    \begin{cases}        
        \,\some{\{X_i(\pids{p}\setminus \pid{p_i}) = P_{\pid{p_i}}\}} &\text{if for each $X(\pids{p}) = C\in \mathscr{C},\, \proj{C}{p_i} \eqr \some{P_{\pid{p_i}}}$ and $\pid{p_i} \in \pids{p}$}\\[-0.5ex]
        \,\none &\text{if for any $X(\pids{p}) = C\in \mathscr{C}$, there exists a $\pid{p} \in \pids{p}$ such that $\proj{C}{p} \eqr \none$}
    \end{cases}
\end{align*}
\vspace{-1.0em}
\caption{Endpoint Projection for Choreographic Procedure Definitions}
\label{epp:c:def:fig}
\end{subfigure}
\label{epp:def:fig}
\vspace{-1.0em}
\caption{Definitions of Endpoint Projection}
\vspace{-1.0em}
\end{figure}

A choreography $C$ is \emph{projectable} onto a process $\pid{r}$ if $\proj{C}{r} \eqr\some{P}$ for some process term $P$. If $\proj{C}{r} \eqr\none$, then $C$ is \emph{not projectable} onto a process $\pid{r}$.
We define EPP of choreographies as $\epp{\cdot} : \setof{Chor} \to \opt{Net}$ in \cref{epp:chor:def:fig}. If $C$ is projectable onto every process in $\pn(C)$, its EPP is a network mapping each such process to its projected process term and all other processes to $\pnil$. Otherwise, the result is $\none$.
In general, $\proj{\cdot}{\pid{p}}$ and $\epp{\cdot}$ are \emph{homomorphisms} of a monoid. This property is crucial for establishing the correctness of EPP concerning KoC, nondeterminism, and out-of-order executions. In \async, we show that $\proj{\cdot}{\pid{p}}$ and $\epp{\cdot}$ are distributive over $\fatsemi$ in \cref{epp:projection:seqcomp}
\begin{proposition}[Distributivity of Process Projection and EPP over $\fatsemi$]\label[proposition]{epp:projection:seqcomp}
    For any choreography $C_1$ and $C_2$, and any process $\pid{p}$:
    $\proj{C_1 \fatsemi C_2}{\pid{p}} = \proj{C_1}{\pid{p}} \fatsemi \proj{C_2}{\pid{p}}$ and $\epp{C_1 \fatsemi C_2} = \epp{C_1} \fatsemi \epp{C_2}$.
\end{proposition}

In \cref{epp:c:def:fig}, we define the EPP for sets of procedure definitions. We overload the notation and write it as $\epp{\mathscr{C}}$, which is a function of the type: $\epp{\cdot} : \mathfrak{C} \to \optg{\mathfrak{P}}$.
A choreographic procedure with $n$ process parameters generates $n$ process procedures, one for the perspective of each parameter. If any procedure body is not projectable onto one of its process parameters, the result is $\none$.

\subsection{Coherence}
\label{epp:coherence}
\begin{figure}[b]
\vspace{-1.0em}
\begin{mathpar}
\thinmuskip=2mu
\thickmuskip=2.5mu
\inferrule*[right={\scriptsize coh-rtcall},sep=0.7em]{\mathscr{C} \propto C \\ \mathscr{C} \propto C' \\ X(\pids{s}) = C'' \in \mathscr{C} \\ |\pids{p}| = |\pids{s}| \\ \proj{C''[\pids{p}/\pids{s}] \fatsemi C'}{q} \eqr\some{P} \\ \proj{C}{q} \eqr\some{Q} \\ \some{P} \mbrsw \some{Q}}
{\mathscr{C} \propto \pid{q} : X(\pids{p}).C' \seq C}
\end{mathpar}
\caption{Coherence of Choreographies (The Rule for Runtime Terms)}
\label{epp:coherence:fig-c}
\vspace{-1.0em}
\end{figure}
We define \emph{coherence} between a choreography and choreographic procedure definitions for handling recursion correctly. It ensures the consistency of projection for partially entered procedure calls. For a runtime call $\pid{q} : X(\pids{p}).C' \seq C$, rule \textsc{coh-rtcall} (\cref{epp:coherence:fig-c}) requires that the projection onto $\pid{q}$ of the unfolded call followed by $C'$ has at least the branches of the projection of $C$ onto $\pid{q}$. The two additional rules for \async, \textsc{coh-rtcom} and \textsc{coh-rtsel}, simply require a coherent continuation (cf. \cref{epp:coherence}).
As with well-formedness, choreographic transitions preserve coherence.
\begin{theorem}[Preservation of Coherence]\label{epp:coherence:preservation:thm}
    For any well-formed choreographic configuration $\langle C, \cstore, \mst, \mathscr{C} \rangle$ in which $C$ and $\mathscr{C}$ are both projectable and $C \propto \mathscr{C}$, if $\langle C, \cstore, \mst, \mathscr{C} \rangle \xrightarrow{\mu} \langle C', \cstore', \mst', \mathscr{C} \rangle$, then the resulting choreography $C'$ is still coherent w.r.t. $\mathscr{C}$, i.e., $C' \propto \mathscr{C}$.
\end{theorem}
\begin{proof}
    By structural induction on the derivation of the transition.
\end{proof}

%% file: sections/epp-correct.tex
\section{Correctness of EPP}
\label{sec:epp:correctness}
We establish the correctness of EPP by showing that choreographies and their projected networks can match each other's transitions. We first prove \emph{completeness} and \emph{soundness} for single-step transitions and then extend both results to multi-step executions.
The main challenge is to track the dependencies that arise when choreographic nondeterminism, asynchrony, out-of-order execution, and general branching interact, so that each alternative behaviour can be matched in the other semantics.
The locality property of \async's semantics enables a compositional approach to proving the correctness of EPP. Each network transition involves a single active process, whose behaviour we analyse separately from the rest of the network. For the unaffected processes, we show that the simulation relation still holds between the resulting network and the projection of the resulting choreography after the matching transitions.

\subsection{Completeness and Soundness of EPP}
\label{sec:epp:completeness:soundness}
The completeness in \cref{epp:completeness:thm} says that all networks resulting for EPP are \emph{complete implementations} of their corresponding choreographies:
for every choreographic transition, there is a corresponding network transition whose resulting network is related by the simulation relation ($\mbrs$) to the projection of the resulting choreography.

The novelty of our reasoning approach is to introduce a \emph{localised semantics} for the process language (\cref{app:local:semantics}). It describes a local transition at an active process $\pid{p}$ in the form $\langle P, \cstore, \mst, \mathscr{P} \rangle \xrightarrow[]{\mu}_{\pid{p}} \langle P', \cstore', \mst', \mathscr{P}\rangle$.
We then establish conversions between localised and network transitions (\cref{net:local:to-network,net:local:of-local,net:local:to-local}). Together with the lax compatibility of merge at the active process (\cref{merge:net:transition}), they allow us to match choreographic transitions with network transitions by analysing the active process's local transition and composing it with the unaffected part of the projected network.
\begin{theorem}[Completeness of EPP]\label{epp:completeness:thm}
   Given a choreographic configuration $\langle C, \cstore, \mst, \mathscr{C} \rangle$ that is well-formed, where $C \propto \mathscr{C}$, $\epp{C} \eqr \some{N}$, and $\epp{\mathscr{C}} \eqr\some{\mathscr{P}}$, for any transition label $\mu$, if $\langle C, \cstore, \mst, \mathscr{C} \rangle \xrightarrow[]{\mu} \langle C', \cstore', \mst', \mathscr{C} \rangle$, then there exist networks $N'$ and $M'$ such that $\langle N, \cstore, \mst, \mathscr{P} \rangle \xrightarrow[]{\mu} \langle N', \cstore', \mst', \mathscr{P} \rangle$, $\epp{C'} \eqr\some{M'}$, and $N' \mbrs M'$.
\end{theorem}
\begin{proof}
   By induction on the transition derivation, using lemmas~\ref{net:local:to-local}, \ref{net:local:to-network} and \cref{merge:net:transition}.
\end{proof}

The soundness in \cref{sec:epp:multi-step} is the converse of the completeness: every network resulting from EPP is a \emph{sound implementation} of its corresponding choreography, and every transition of $N$ is prescribed by $C$, with the resulting network retaining the simulation relation.
We give the formal definition in \cref{epp:soundness:thm}.
We require $N \mbrs M$ for the generalisation to multi-step transitions (\cref{sec:epp:multi-step}).
Completeness starts from a choreographic transition whose message queue premises already hold, while the soundness proof need to use message queue consistency to ensure that a network receive matches a pending receive in the choreography. To structure the queue consistency argument, we define the transition relation $\mst \xrightarrow{\mu} \mst'$ for messaging states in \cref{app:mst:semantics}. We can then replace the messaging states of a transition when applying the induction hypothesis to the continuation. We use \cref{mst:semantics:remove-passive-head} to isolate a messaging state transition from a network transition.
\begin{theorem}[Soundness of EPP]\label{epp:soundness:thm}
    Given a choreographic configuration $\langle C, \cstore, \mst, \mathscr{C} \rangle$ that is well-formed, where $C \propto \mathscr{C}$, $\epp{C} \eqr \some{M}$, and $\epp{\mathscr{C}} \eqr\some{\mathscr{P}}$, for any networks $N$, $N'$, and transition label $\mu$ where $N \mbrs M$, if $\langle N, \cstore, \mst, \mathscr{P} \rangle \xrightarrow[]{\mu} \langle N', \cstore', \mst', \mathscr{P} \rangle$, then there exists a choreography $C'$ and a network $M'$ such that $\langle C, \cstore, \mst, \mathscr{C} \rangle \xrightarrow[]{\mu} \langle C', \cstore', \mst', \mathscr{C} \rangle$, $\epp{C'}\eqr\some{M'}$, and $N' \mbrs M'$.
\end{theorem}
\begin{proof}
    By induction on the choreography $C$. By \cref{net:local:to-local}, we identify the active process and use its local transition to derive the matching choreographic step. For steps delayed past a pending receive, \cref{net:semantics:queue-step,mst:semantics:remove-passive-head,chor:semantics:queue-congr} justify applying the induction hypothesis with the continuation's messaging states and then restoring the original messaging states. For delayed conditionals and choices, \cref{epp:projection:seqcomp,chor:semantics:seqcomp-border} relate the branch transitions to their continuation.
\end{proof}

\subsection{Correctness of Multi-Step Executions}
\label{sec:epp:multi-step}
We use a standard transitive closure to define the multi-step transitions for both choreographies and networks, written as $\langle C, \cstore, \mst, \mathscr{C} \rangle \xtwoheadrightarrow{\,\tls\,} \langle C', \cstore', \mst', \mathscr{C} \rangle$ and $\langle N, \cstore, \mst, \mathscr{P} \rangle \xtwoheadrightarrow{\,\tls\,} \langle N', \cstore', \mst', \mathscr{P} \rangle$.
The sequence $\tls$ is a list of transition labels that can be empty.
The detailed definitions are shown in \cref{app:multi}.
We show the correctness of EPP for multi-step executions by the completeness \cref{epp:multi-step-completeness:thm} and soundness \cref{epp:multi-step-soundness:thm}:
\begin{theorem}[Completeness of Multi-Step Executions]\label{epp:multi-step-completeness:thm}
    Let $\langle C, \cstore, \mst, \mathscr{C} \rangle$ be a well-formed choreographic configuration, where $C \propto \mathscr{C}$, $\epp{C} \eqr \some{N}$, and $\epp{\mathscr{C}} \eqr\some{\mathscr{P}}$, for any list of transition labels $\tls$, if $\langle C, \cstore, \mst, \mathscr{C} \rangle \xtwoheadrightarrow{\,\tls\,} \langle C', \cstore', \mst', \mathscr{C} \rangle$, then there exists networks $N'$ and $M'$ such that $\langle N, \cstore, \mst, \mathscr{P} \rangle \xtwoheadrightarrow{\,\tls\,} \langle N', \cstore', \mst', \mathscr{P} \rangle$, $\epp{C'} \eqr\some{M'}$, and $N' \mbrs M'$.
\end{theorem}

\begin{theorem}[Soundness of Multi-Step Executions]\label{epp:multi-step-soundness:thm}
    Let $\langle C, \cstore, \mst, \mathscr{C} \rangle$ be a well-formed choreographic configuration with $C \propto \mathscr{C}$, $\epp{C} \eqr \some{M}$, and $\epp{\mathscr{C}} \eqr \some{\mathscr{P}}$. For any networks $N$, $N'$ and any list $\tls$ of transition labels, if $N \mbrs M$ and $\langle N, \cstore, \mst, \mathscr{P} \rangle \xtwoheadrightarrow{\,\tls\,} \langle N', \cstore', \mst', \mathscr{P} \rangle$, then there exist a choreography $C'$ and a network $M'$ such that $\langle C, \cstore, \mst, \mathscr{C} \rangle \xtwoheadrightarrow{\,\tls\,} \langle C', \cstore', \mst', \mathscr{C} \rangle$,$\epp{C'} \eqr \some{M'}$, and $N' \mbrs M'$.
\end{theorem}
We prove both theorems by induction on the derivation of the multi-step transition. We use the completeness and soundness~\cref{epp:completeness:thm,epp:soundness:thm}, the preservation of well-formedness (\cref{chor:wf:preservation:thm}), coherence (\cref{epp:coherence:preservation:thm}), and simulation relation (\cref{merge:mbrs:net:transition}).

%% file: sections/comm-safe-deadlock-free.tex
\section{The Correctness of EPP Guarantees Safety and Liveness Properties}
\label{sec:epp:properties}
The correctness of EPP guarantees important safety and liveness properties of networks projected from choreographies. We establish \emph{communication safety} in \cref{epp:comm-safety} and \emph{deadlock freedom} in \cref{epp:deadlock-freedom}.
In an asynchronous system, the sender can proceed once it places a message in the receiver's queue, while the corresponding receive remains pending. Establishing communication safety and deadlock freedom is thus more involved than in the synchronous setting, since we must also reason about which messages are queued and when they can be received.

\subsection{Communication Safety}
\label{epp:comm-safety}
Communication safety is an important safety property of communicating processes. It ensures the absence of mismatched communications. A mismatch occurs, for instance, when a receiver expects a value but the message to be received is a selection label.

\Cref{comm-act:fig} defines communication actions and their duals, and \cref{enabled-comm-act:fig} gives the enabled communication actions of a process term for a given process. We characterise communication-safe states in \cref{def:comm-safety-state} by checking whether the sender's and receiver's enabled actions are compatible given the contents of their message queue. We define this compatibility in \cref{def:queue-compatibility}.
\begin{figure}[b]
\vspace{-1.0em}
\begin{align*}
   \text{Communication Action (\setof{Act})}\quad \alpha \ \metaDeff \ ! \cmid \ ? \cmid \oplus_{\albl} \cmid \&_{\albl}
\end{align*}
\vspace{-1.5em}
\begin{align*}
\begin{aligned}[c]
\overline{!} \triangleq \ ?
\end{aligned}
\quad\quad
\begin{aligned}[c]
\overline{?} \triangleq \ !
\end{aligned}
\quad\quad
\begin{aligned}[c]
\overline{\oplus_{\albl}} \triangleq \&_{\albl}
\end{aligned}
\quad\quad
\begin{aligned}[c]
\overline{\&_{\albl}} \triangleq \oplus_{\albl}
\end{aligned}
\end{align*}
\vspace{-1.0em}
\caption{Communication Actions and Their Duals}
\label{comm-act:fig}
\vspace{-1.0em}
\end{figure}

\begin{figure}
\vspace{-1.0em}
\begin{align*}
\begin{aligned}[c]
    \pnil \octopusop \pid{p} &\triangleq \emptyset
\end{aligned}
\quad\quad
\begin{aligned}[c]
    x \metaDef e \seq P \octopusop \pid{p} &\triangleq \emptyset
\end{aligned}
\quad\quad
\begin{aligned}[c]
    X(\pids{q}) \seq P \octopusop \pid{p} &\triangleq \emptyset
\end{aligned}
\end{align*}
\vspace{-1.7em}
\begin{align*}
\begin{aligned}[c]
    \condif e\condthen P_1\condelse P_2 \seq P \octopusop \pid{p} &\triangleq \emptyset
\end{aligned}
\quad\quad
\begin{aligned}[c]
    P_1 \choice P_2 \seq P \octopusop \pid{p} &\triangleq P_1 \octopusop \pid{p} \cup P_2 \octopusop \pid{p}
\end{aligned}
\end{align*}
\vspace{-1.7em}
\begin{align*}
\begin{aligned}[c]
    \psend{q}{e}\seq P \octopusop \pid{p} &\triangleq
    \begin{cases}
        \ \{!\} &\text{if $\pid{p} = \pid{q}$}\\[-0.5ex]
        \ \emptyset &\text{otherwise}
    \end{cases}\\[-0.5ex]
    \psel{q}\seq P \octopusop \pid{p} &\triangleq
    \begin{cases}
        \ \{\oplus_{\albl}\} &\text{if $\pid{p} = \pid{q}$}\\[-0.5ex]
        \ \emptyset &\text{otherwise}
    \end{cases}
\end{aligned}
\quad\quad
\begin{aligned}[c]
    \precv{q}{x}\seq P \octopusop \pid{p} &\triangleq
    \begin{cases}
        \ \{?\} &\text{if $\pid{p} = \pid{q}$}\\[-0.5ex]
        \ \emptyset &\text{otherwise}
    \end{cases}\\[-0.5ex]
    \precvl{q} \octopusop \pid{p} &\triangleq
    \begin{cases}
        \ \{\&_{\albl_i}\}_{i\in I} &\text{if $\pid{p} = \pid{q}$}\\[-0.5ex]
        \ \emptyset &\text{otherwise}
    \end{cases}
\end{aligned}
\end{align*}
\caption{Enabled Communication Actions}
\label{enabled-comm-act:fig}
\vspace{-1.0em}
\end{figure}

\begin{definition}[Message Queue Compatibility]\label[definition]{def:queue-compatibility}
    Consider the message queue and two sets of communication actions, one for the receiver and one for the sender.
    When the queue is non-empty, the sets are compatible if and only if the receiver's set is input compatible with the pending receive for the message at the head of the queue. The pending receive is $?$ for a value and $\&_{\albl}$ for a selection label $\albl$.
    When the queue is empty, the sets are compatible if and only if the receiver's set is input compatible with $\overline{\alpha}$ for every value send or label selection $\alpha$ in the sender's set.
\end{definition}

\begin{definition}[Communication-Safe State]\label[definition]{def:comm-safety-state}
    A network $N$ is \emph{communication safe} at a messaging state $\mst$ if and only if, for every ordered pair of distinct processes $\pid{p}$ and $\pid{q}$, the enabled actions $N(\pid{q}) \octopusop \pid{p}$ of the receiver and $N(\pid{p}) \octopusop \pid{q}$ of the sender are compatible for the message queue $\mst(\pid{q},\pid{p})$.
    A network configuration $\langle N,\cstore,\mst,\mathscr{P}\rangle$ is communication safe if and only if $N$ is communication safe at $\mst$.
\end{definition}

\begin{theorem}[Communication Safety]\label{epp:comm-safety:thm}
    Let $\langle C, \cstore, \mst, \mathscr{C} \rangle$ be a well-formed choreographic configuration, where $C \propto \mathscr{C}$, $\epp{C} \eqr \some{N}$, and $\epp{\mathscr{C}} \eqr \some{\mathscr{P}}$. If $\langle N, \cstore, \mst, \mathscr{P} \rangle \xtwoheadrightarrow{\,\tls\,} \langle N', \cstore', \mst', \mathscr{P} \rangle$, then $\langle N', \cstore', \mst', \mathscr{P} \rangle$ is communication safe.
\end{theorem}
\begin{proof}
    By \cref{epp:multi-step-soundness:thm}, the network execution is matched by
    $\langle C,\cstore,\mst,\mathscr{C}\rangle \xtwoheadrightarrow{\,\tls\,}
    \langle C',\cstore',\mst',\mathscr{C}\rangle$ with $\epp{C'}=\;\some{M'}$ and $N'\mbrs M'$.
    By preservation of well-formedness and coherence (\cref{chor:wf:preservation:thm,epp:coherence:preservation:thm}), \cref{epp:comm-safe-state} gives communication safety of $M'$ at $\mst'$ and non-empty branch lists throughout $M'$.
    By \cref{epp:comm-safe-state:more-branches}, $N'$ is communication safe at $\mst'$.
\end{proof}

\subsection{Deadlock Freedom}
\label{epp:deadlock-freedom}
A deadlock occurs when a network reaches a configuration with no outgoing transition before it terminates. We establish deadlock freedom for networks projected from well-formed, coherent choreographies. After any sequence of transitions from such a projection, the reached configuration can take another step unless its network is the terminated network $\nnil$.

In \async, the progress proof relies on messaging state consistency. In the case of choreographic choice, the messaging state must be consistent with each branch followed by the common continuation. \Cref{epp:progress:thm} establishes progress for projectable choreographies, and \cref{epp:deadlock-freedom:thm} uses the correctness of EPP to establish deadlock freedom for their projections.

\begin{theorem}[Progress of Projectable Choreographies]\label{epp:progress:thm}
    Given a choreographic configuration $\langle C, \cstore, \mst, \mathscr{C} \rangle$ that is well-formed, where $C \propto \mathscr{C}$ and both $C$ and $\mathscr{C}$ are projectable, if $C \neq \cnil$, then there exist a choreography $C'$, a global store $\cstore'$, a messaging state $\mst'$, and a transition label $\mu$ such that $\langle C, \cstore, \mst, \mathscr{C} \rangle \xrightarrow{\mu} \langle C', \cstore', \mst', \mathscr{C} \rangle$.
\end{theorem}
\begin{proof}
    By induction on the size of $C$, strengthening the induction hypothesis to allow a projectable continuation $D$ with $\mst \cst (C \fatsemi D)$.
\end{proof}

\begin{theorem}[Deadlock Freedom]\label{epp:deadlock-freedom:thm}
    Given a choreographic configuration $\langle C, \cstore, \mst, \mathscr{C} \rangle$ that is well-formed, where $C \propto \mathscr{C}$, $\epp{C} \eqr \some{N}$, and $\epp{\mathscr{C}} \eqr \some{\mathscr{P}}$, if $\langle N, \cstore, \mst, \mathscr{P} \rangle \xtwoheadrightarrow{\,\tls\,} \langle N', \cstore', \mst', \mathscr{P} \rangle$ and $N' \neq \nnil$, then there exist a network $N''$, a global store $\cstore''$, a messaging state $\mst''$, and a transition label $\mu$ such that $\langle N', \cstore', \mst', \mathscr{P} \rangle \xrightarrow{\mu} \langle N'', \cstore'', \mst'', \mathscr{P} \rangle$.
\end{theorem}
\begin{proof}
    By \cref{epp:multi-step-soundness:thm}, the choreography reaches $\langle C', \cstore', \mst', \mathscr{C} \rangle$ along $\tls$, with $\epp{C'} \eqr \some{M'}$ and $N' \mbrs M'$. Preservation of well-formedness and coherence (\cref{chor:wf:preservation:thm,epp:coherence:preservation:thm}) applies along this execution. \Cref{epp:progress:thm} gives a step from $\langle C', \cstore', \mst', \mathscr{C} \rangle$. By \cref{epp:completeness:thm}, $M'$ can take that step from $\cstore'$ and $\mst'$. By \cref{merge:mbrs:net:transition}, $N'$ can take the same step from $\cstore'$ and $\mst'$.
\end{proof}

%% file: sections/related-work.tex
\section{Related Work}
\label{sec:related-work}
\subsubsection*{Asynchrony and Nondeterminism in CP}
Ozone~\cite{PPM24} studies fully out-of-order choreographies. In addition to allowing the sender to continue while a message is pending, Ozone also allows a process waiting for one message to receive
another and use it in subsequent computations, provided those actions do not depend on the pending receive. In \async, messages from a given sender to a given receiver are received in
FIFO order.
However, we are able to encode Ozone's asynchronous receive feature with choreographic choice by writing a branch for each desired schedule~\cite{PPM24}. Ozone does not support choreographic choice, which can express behaviours outside Ozone, such as binding messages to
different variables depending on their arrival order. The authors implement their approach as a library for Choral~\cite{GMP24}, while we mechanise \async{} in Lean~4.

In Minimal Choreographies, \citet{CM17a} give separate rules for sending and receiving and prove operational correspondence with networks that use FIFO queues. Kalas~\cite{PGSN22} also supports asynchronous communication. Its choreographic semantics allows a sender to proceed before its message is received, as in \citet{CM13}. For knowledge of choice, EPP in Kalas uses initial Boolean selections to combine alternatives at the top level. \async{} adopts \mech{}'s algebraic approach to full merge. Although these theories do not model choreographic nondeterminism, their studies of asynchrony are inspiring.

For theories that model choreographic nondeterminism, the textbook~\cite{M23} outlines choreographic choice and asynchronous communication as separate extensions. \mech{}~\cite{QPM26} gives the first correct semantics for choreographic choice with synchronous communication and proves the correctness of EPP. As discussed in \cref{sec:nondeterminism-async}, a single communication in \mech can resolve choices at both endpoints.
In \async{}, the send and receive occur in separate transitions, and each can resolve choices at its active process.

\subsubsection*{Asynchrony and Mixed Choice in MPST}
Multiparty session types (MPST) use a global type to describe communication in a protocol and project that type to a local type for each participant. In the original theory of asynchronous MPST~\cite{HYC08,HYC16}, a designated sender selects a branch by sending a label to a designated receiver. In the asynchronous system of \citet{CDP12}, an external choice can offer receives from different
senders. Each local choice contains only inputs or only outputs.
Zooid~\cite{CGY21} mechanises asynchronous MPST in Rocq and proves trace equivalence between global types and systems of projected local types.

\citet{PY24a} study mixed choice in synchronous multiparty sessions. A single choice can contain both sends and receives, possibly involving different peers. A matching send and receive resolve the choices at both endpoints in one synchronised transition. They show that separate choices across peers already express the $M$ synchronisation pattern and that mixed choices across peers also express the star pattern~\cite{GGS08,PNG13}.

\citet{BD26} adapt the mixed choice calculus of \citet{PY24a} to asynchronous communication. They type sessions directly against global types and prove subject reduction and session fidelity.
\citet{BHVT26} propose asymmetric mixed choice for asynchronous MPST. Other participants can act in the left branch before a designated observer commits by receiving on the left or sending on the right. The runtime can purge stale messages from the abandoned branch.
They prove weak operational correspondence for endpoint projection, treating choice instantiation and stale message purging as silent steps, highlighting the challenges of combining asynchronous communication with protocol-level nondeterminism.

\subsubsection*{From Mechanised Theory to Certified Compilation for CP}
Including aforementioned Kalas and Mech, several studies have explored building certified compiler frameworks for CP with existing mechanised theories. Kalas formalises its theory in HOL4 and passes translated endpoint programs to the verified CakeML compiler, which generates executable machine code.
\citet{CMP21b} study endpoint projection for Core Choreographies, a language with synchronous communication, binary label selections, conditionals, and tail recursion.
They mechanise the projection to Stateful Processes in Rocq and prove operational correspondence up to pruning. \citet{CLM23} extract this projection to Haskell and use it in hacc, which generates one executable Jolie service per projected process. The certification covers endpoint projection, with the Haskell wrapper and Jolie code generator left to manual inspection.
As discussed in \cref{sec:introduction}, Pirouette~\cite{HG22} and $\lambda_{\mathrm{QC}}$~\cite{SHC25} are both well-known higher-order functional choreographic languages, mechanised in Rocq. However, they can express very limited concurrency features since they both support synchronous communication and require global synchronisation, even for non-involved processes, at every choreography call.
\async{} supports asynchronous communication and choreographic choice and allows processes to enter procedure calls independently.
Its mechanised EPP correctness certifies the translation from choreographies to networks.
We plan to use \async{} as an intermediate language in a future certified CP compilation framework. Closure conversion~\cite{MMH96} offers a possible starting point for compiling higher-order fragments of Pirouette and $\lambda_{\mathrm{QC}}$
into \mech{}.

%% file: sections/conclusion-future-work.tex
\section{Conclusion and Future Work}
\label{sec:conclusion:future-work}
We have presented \async, a new CP framework that combines asynchronous communication and choreographic nondeterminism. It supports general branching, out-of-order execution, and general recursion with independent process entry.
We proved the correctness of EPP and established communication safety and deadlock freedom. We mechanised the entire theory in Lean~4.
The locality property of \async's semantics enables compositional reasoning and gives the proofs a clearer structure. This is reflected in the size of the Lean~4 mechanisations, with only around 20,000 lines of code for \async and around 40,000 for \mech, which supports fewer features.

As discussed in \cref{sec:related-work}, researchers have investigated developing certified compilation frameworks for CP based on existing mechanised theories. More recently, researchers also starting building domain-specific languages for stateful process and implementing security protocols in distributed systems in Lean 4 with its rich metaprogramming features in CSLib~\cite{cslib2026}.
Therefore, we plan to further investigate how to build certified and semi-certified compiler that connect the rich theories of CP to mainstream tooling. We expect \async to serve as an intermediate language for such a framework. 
\async's local language agnostic design gives us flexibility in choosing target languages for code generation.

\section*{AI Use}
The authors have used Codex (models Sol 6.1 and Astra 6) for facilitating the mechanisation. Our uses include proof refactoring and optimisation (reducing heartbeats), proof search, and creating documentation for proofs. AI has not been involved in designing any theory presented in the work or defined/declared in the Lean 4 mechanisation.

\section*{Data-Availability Statement}

The source code for the complete mechanisation of our theory in the Lean~4 proof assistant is included in the supplementary material accompanying this submission, in the folder ``nyarlathotep''. Build instructions and a detailed explanation of the correspondence between the paper formalisation and the mechanisation are provided in the file ``README.md''. 
We intend to submit this artifact for Artifact Evaluation.

%% file: appendix/pn.tex
\section{Process Names}
\label[appendix]{app:pn}

\begin{definition}[Process Names in Choreographies and Instructions]\label[definition]{chor:syntax:pn}
\begin{align*}
\begin{gathered}
\pn(\cnil) \triangleq \emptyset
\quad
\pn(I \seq C) \triangleq \pn(I) \cup \pn(C)
\quad
\pn(\pid{p}.x \metaDef e) \triangleq \{\pid{p}\}
\\
\pn(\gencom) \triangleq \{\pid{p}, \pid{q}\}
\quad
\pn(\gensel) \triangleq \{\pid{p}, \pid{q}\}
\quad
\mathshade{\pn(\genrtcom) \triangleq \{\pid{q}\}}
\quad
\mathshade{\pn(\genrtsel) \triangleq \{\pid{q}\}}
\\
\pn(\condif \pid{p}.e \condthen C_1 \condelse C_2) \triangleq \{\pid{p}\} \cup \pn(C_1) \cup \pn(C_2)
\quad
\pn(C_1 \choice_\pid{p} C_2) \triangleq \{ \pid{p} \} \cup \pn(C_1) \cup \pn(C_2)
\\
\pn(X(\pids{p})) \triangleq \{\pids{p}\}
\quad
\pn(\pid{q} : X(\pids{p}).C ) \triangleq \{\pid{q}\}
\end{gathered}
\end{align*}
\end{definition}

\begin{definition}[Process Names in Transition Labels]\label[definition]{chor:semantics:pntl}
\begin{align*}
\begin{gathered}
\pn(\tlskip{p}) \triangleq \{ \pid{p} \}
\quad
\pn(\tlassign{p}) \triangleq \{ \pid{p} \}
\quad
\pn(\tlcall{p}) \triangleq \{ \pid{p} \}\\
\mathshade{\pn(\tlcoms{p}{v}{q}) \triangleq \{\pid{p}\}}
\quad
\mathshade{\pn(\tlcomr{p}{v}{q}) \triangleq \{\pid{q}\}}
\quad
\mathshade{\pn(\tlsels{p}{q}{l}) \triangleq \{\pid{p}\}}
\quad
\mathshade{\pn(\tlselr{p}{q}{l}) \triangleq \{\pid{q}\}}\\
\pn(\tlthen{p}) \triangleq \{ \pid{p} \}
\quad
\pn(\tlelse{p}) \triangleq \{ \pid{p} \}\\
\mathshade{\pn(\tlleft{p}{\mu}) \triangleq \{ \pid{p} \} \cup \pn(\mu)}
\quad
\mathshade{\pn(\tlright{p}{\mu}) \triangleq \{ \pid{p} \} \cup \pn(\mu)}
\end{gathered}
\end{align*}
\end{definition}

%% file: appendix/store-and-subst.tex
\section{Stores and Process Name Substitutions}
\label{app:stores-subst}
\begin{definition}[Update Operations for  Stores]\label[definition]{app:def:store-update}
\begin{align*}
\begin{aligned}[c]
\sigma[y \mapsto v] (x) &\triangleq
\begin{cases}
\ v &\text{if $x = y$}\\
\ \sigma(x) &\text{otherwise}
\end{cases}
\end{aligned}
\quad\quad\quad
\begin{aligned}[c]
\cstore[\pid{q}.x \mapsto v] (\pid{p}) &\triangleq
\begin{cases}
\ \cstore(\pid{p})[x \mapsto v] &\text{if $\pid{p} = \pid{q}$}\\
\ \cstore(\pid{p}) &\text{otherwise}
\end{cases}
\end{aligned}
\end{align*}
\end{definition}

\begin{definition}[Name Substitution for Process Names and Sequences of Process Names]\label[definition]{chor:semantics:pid-pids-substitution}
\begin{align*}
\begin{aligned}[c]
    \pid{p}[\pid{r}/\pid{s}] &\triangleq
    \begin{cases}
    \,\pid{r} &\text{if $\pid{p} = \pid{s}$}\\
    \,\pid{p} &\text{otherwise}
    \end{cases}
\end{aligned}
\quad\quad\quad\quad
\begin{aligned}[c]
    (\pid{p}_1, \dots \pid{p}_n) [\pid{r}/\pid{s}] &\triangleq (\pid{p}_1[\pid{r}/\pid{s}], \dots, \pid{p}_n[\pid{r}/\pid{s}])
\end{aligned}
\end{align*}
\end{definition}

\begin{definition}[Name Substitution for Choreographies and Instructions]\label[definition]{chor:semantics:chor-substitution}
\begin{align*}
\begin{aligned}[c]
\cnil[\pid{r}/\pid{s}] &\triangleq \cnil\\
(\pid{p}.x \metaDef e)[\pid{r}/\pid{s}] &\triangleq (\pid{p}[\pid{r}/\pid{s}]).x \metaDef e\\
(\pid{q} : X(\pids{p}).C) [\pid{r}/\pid{s}] &\triangleq \pid{q}[\pid{r}/\pid{s}] : X(\pids{p}[\pid{r}/\pid{s}]).(C[\pid{r}/\pid{s}])\\
(\gencom)[\pid{r}/\pid{s}] &\triangleq (\pid{p}[\pid{r}/\pid{s}]).e \tto (\pid{q}[\pid{r}/\pid{s}]).x\\
(\genrtcom)[\pid{r}/\pid{s}] &\triangleq (\pid{p}[\pid{r}/\pid{s}]).e \rto (\pid{q}[\pid{r}/\pid{s}]).x
\end{aligned}
\quad\quad
\begin{aligned}[c]
(I\seq C)[\pid{r}/\pid{s}] &\triangleq (I[\pid{r}/\pid{s}])\seq (C[\pid{r}/\pid{s}])\\
(X(\pids{p})[\pid{r}/\pid{s}]) &\triangleq X(\pids{p}[\pid{r}/\pid{s}])\\
(C_1 \choice_\pid{p} C_2) [\pid{r}/\pid{s}] &\triangleq (C_1[\pid{r}/\pid{s}]) \choice_{\pid{p}[\pid{r}/\pid{s}]}(C_2[\pid{r}/\pid{s}])\\
(\gensel)[\pid{r}/\pid{s}] &\triangleq (\pid{p}[\pid{r}/\pid{s}]) \tto (\pid{q}[\pid{r}/\pid{s}])[\albl]\\
(\genrtsel)[\pid{r}/\pid{s}] &\triangleq (\pid{p}[\pid{r}/\pid{s}]) \rto (\pid{q}[\pid{r}/\pid{s}])[\albl]
\end{aligned}
\end{align*}
\vspace{-1.0em}
\begin{align*}
(\condif \pid{p}.e\condthen C_1\condelse C_2) [\pid{r}/\pid{s}] &\triangleq \condif(\pid{p}[\pid{r}/\pid{s}]).e \condthen(C_1[\pid{r}/\pid{s}]) \condelse (C_2[\pid{r}/\pid{s}])\\
C[\pid{r}_1, \dots, \pid{r}_n/\pid{s}_1, \dots, \pid{s}_n] &\triangleq C[\pid{r}_1/\pid{s}_1]\dots[\pid{r}_n/\pid{s}_n]
\end{align*}
\end{definition}

\begin{definition}[Name Substitution for Process Terms and Instructions]\label[definition]{proc:semantics:proc-substitution}
\begin{align*}
\begin{aligned}[c]
\pnil[\pid{r}/\pid{s}] &\triangleq \pnil\\
(\precvl p)[\pid{r}/\pid{s}] &\triangleq \precvl p[\pid{r}/\pid{s}]\\
(\psend{p}{e})[\pid{r}/\pid{s}] &\triangleq \psend{p[\pid{r}/\pid{s}]}{e}\\
(x \metaDef e)[\pid{r}/\pid{s}] &\triangleq x \metaDef e
\end{aligned}
\quad\quad
\begin{aligned}[c]
(I\seq P)[\pid{r}/\pid{s}] &\triangleq (I[\pid{r}/\pid{s}])\seq (P[\pid{r}/\pid{s}])\\
(\psel{p})[\pid{r}/\pid{s}] &\triangleq \psel{p[\pid{r}/\pid{s}]}\\
(\precv{p}{x})[\pid{r}/\pid{s}] &\triangleq \precv{p[\pid{r}/\pid{s}]}{x}\\
(X(\pids{p})[\pid{r}/\pid{s}]) &\triangleq X(\pids{p}[\pid{r}/\pid{s}])
\end{aligned}
\end{align*}
\vspace{-1.0em}
\begin{align*}
(\condif e\condthen P_1\condelse P_2) [\pid{r}/\pid{s}] &\triangleq \condif e \condthen(P_1[\pid{r}/\pid{s}]) \condelse (P_2[\pid{r}/\pid{s}])\\
(P_1 \choice P_2) [\pid{r}/\pid{s}] &\triangleq (P _1[\pid{r}/\pid{s}]) \choice (P_2[\pid{r}/\pid{s}])\\
P[\pid{r}_1, \dots, \pid{r}_n/\pid{s}_1, \dots, \pid{s}_n] &\triangleq P[\pid{r}_1/\pid{s}_1]\dots[\pid{r}_n/\pid{s}_n]
\end{align*}
\end{definition}

%% file: appendix/wf.tex
\section{Well-formedness of Choreographies}
\label[appendix]{app:chor:wf}
We define the \emph{well-formedness of choreographies} with the notation $\langle \mathscr{C}, \{\pids{r}\}, C \rangle \checkmark$.
It denotes that a choreography $C$ is well-formed under a set of choreographic procedures $\mathscr{C}$ and a set $\{\pids{r}\}$ of processes that may be involved.
Figure~\ref{chor:wf:fig} gives the formal definition of well-formed choreographies.

\begin{figure}
\begin{mathpar}
\inferrule*[right={\scriptsize wf-end}]{ }
{ \langle \mathscr{C}, \{\pids{r}\}, \termination \rangle \checkmark}

\inferrule*[right={\scriptsize wf-assign}]{\pid{p} \in \{\pids{r}\} \\ \langle \mathscr{C}, \{\pids{r}\}, C \rangle \checkmark} 
{ \langle \mathscr{C}, \{\pids{r}\}, \pid{p}.x \metaDef e \seq C \rangle \checkmark }

\inferrule*[right={\scriptsize wf-com}]{\pid{p} \neq \pid{q} \\ \pid{p}, \pid{q} \in \{\pids{r}\} \\ \langle \mathscr{C}, \{\pids{r}\}, C \rangle \checkmark}  
{ \langle \mathscr{C}, \{\pids{r}\}, \gencom \seq C \rangle \checkmark}

\mathshade{
\inferrule*[right={\scriptsize wf-rtcom}]{\pid{p} \neq \pid{q} \\ \pid{q} \in \{\pids{r}\} \\ \langle \mathscr{C}, \{\pids{r}\}, C \rangle \checkmark}  
{ \langle \mathscr{C}, \{\pids{r}\}, \genrtcom \seq C \rangle \checkmark}
}

\inferrule*[right={\scriptsize wf-sel}]{\pid{p} \neq \pid{q} \\ \pid{p}, \pid{q} \in \{\pids{r}\} \\ \langle \mathscr{C}, \{\pids{r}\}, C \rangle \checkmark }
{ \langle \mathscr{C}, \{\pids{r}\}, \gensel \seq C \rangle \checkmark }

\mathshade{
\inferrule*[right={\scriptsize wf-rtsel}]{\pid{p} \neq \pid{q} \\ \pid{q} \in \{\pids{r}\} \\ \langle \mathscr{C}, \{\pids{r}\}, C \rangle \checkmark }
{ \langle \mathscr{C}, \{\pids{r}\}, \genrtsel \seq C \rangle \checkmark }
}

\inferrule*[right={\scriptsize wf-cond}]{ \pid{p} \in \{ \pids{r}\} \\ \langle \mathscr{C}, \{\pids{r}\}, C_1 \rangle \checkmark \\ \langle \mathscr{C}, \{\pids{r}\}, C_2 \rangle \checkmark \\ \langle \mathscr{C}, \{\pids{r}\}, C \rangle \checkmark}
{\langle \mathscr{C}, \{\pids{r}\}, \condif \pid{p}.e\condthen C_1\condelse C_2 \seq C \rangle \checkmark}

\inferrule*[right={\scriptsize wf-choice}]{ \pid{p} \in \{\pids{r}\} \\ \langle \mathscr{C}, \{\pids{r}\}, C_1 \rangle \checkmark \\ \langle \mathscr{C}, \{\pids{r}\}, C_2 \rangle \checkmark \\ \langle \mathscr{C}, \{\pids{r}\}, C \rangle \checkmark}
{\langle \mathscr{C}, \{\pids{r}\}, C_1 \choice_\pid{p} C_2 \seq C \rangle \checkmark}

\inferrule*[right={\scriptsize wf-call}]{ \distinct(\pids{p}) \\ \{\pids{p}\} \subseteq \{\pids{r}\} \\ X(\pids{q}) = C' \in \mathscr{C} \\ |\pids{p}| = |\pids{q}| \\ \langle \mathscr{C}, \{\pids{r}\}, C \rangle \checkmark }
{\langle \mathscr{C}, \{\pids{r}\}, X(\pids{p}) \seq C  \rangle \checkmark}

\inferrule*[right={\scriptsize wf-rt-call}]{ \distinct(\pids{p}) \quad \pid{q} \in \{\pids{r}\} \quad \pid{q} \in \{\pids{p}\} \quad X(\pids{s}) = C'' \in \mathscr{C} \quad |\pids{p}| = |\pids{s}| \quad \langle \mathscr{C}, \{\pids{r}\}, C \rangle \checkmark}
{ \langle \mathscr{C}, \{\pids{r}\}, \pid{q} : X(\pids{p}).C' \seq C \rangle \checkmark }
\end{mathpar}
\caption{Well-Formedness of Choreographies}
\label{chor:wf:fig}
\end{figure}

%% file: appendix/merge.tex
\section{Merge and the Branching Order}
\label[appendix]{app:merge}
We take an algebraic approach for defining the merge operation $\sqcup$, forming \emph{join-semilattice} structures $(\opt{Process}, \sqcup)$ and $(\opt{PInstr}, \sqcup)$. The type signatures of ($\sqcup$) are:
\begin{spreadlines}{0pt}
\begin{align*}
\begin{aligned}[c]
    \sqcup : \opt{Process}\rightarrow \opt{Process} \rightarrow \opt{Process}
\end{aligned}
\quad\quad\quad
\begin{aligned}[c]
    \sqcup : \opt{PInstr}\rightarrow \opt{PInstr} \rightarrow \opt{PInstr}
\end{aligned}
\end{align*}
\end{spreadlines}
\\[-4.0ex]
We formally define $(\sqcup)$ for process terms and instructions in \cref{merge:proc:fig} and \cref{merge:instr:fig}. Note that $\opterm{P}$ is a term of the option type $\opt{Process}$, i.e., either $\some{P}$ or a none term $\none$, and $\opterm{I}$ is a term of $\opt{PInstr}$.
\begin{figure}
\vspace{-1.0em}
\begin{subfigure}{1.0\textwidth}
\begin{spreadlines}{0pt}
\begin{align*}
\begin{aligned}[c]
    \some{\pnil} \sqcup \some{\pnil}\ \triangleq\ \some{\pnil}
\end{aligned}
\quad
\begin{aligned}[c]
    \some{I_1 \seq P_1} \sqcup \some{I_2 \seq P_2} \ \triangleq
    \begin{cases}
    \;\some{I \seq P} &\text{if $\some{I_1} \sqcup \some{I_2} \eqw \some{I}$ and $\some{P_1} \sqcup \some{P_2} \eqw\some{P}$} \\[-0.5ex]
    \;\none &\text{otherwise}
    \end{cases}
\end{aligned}
\end{align*}
\vspace{-1.0em}
\end{spreadlines}
\begin{spreadlines}{0pt}
\begin{align*}
    &\some{\precvl {p_1}} \sqcup \some{\precvlj {p_2}} \ \triangleq\\
    &\begin{cases}
    \some{\pid{p_1}\&\{\albl_k \!:\! R_k\}_{k\in \idx{I} \cap \idx{J}} \!\cup\! \{\albl_i \!:\! P_i\}_{i\in \idx{I} \setminus \idx{J}} \!\cup\! \{\albl_j \!:\! Q_j\}_{j\in \idx{J} \setminus \idx{I}}}  &\quad{\begin{aligned}[t]&\text{if $\pid{p_1}\!=\!\pid{p_2}$, and for each $k\!\in\! \idx{I} \!\cap\! \idx{J},$}\\&\text{$\some{P_k} \!\sqcup\! \some{Q_k} \eqw \some{R_k}$}\end{aligned}} \\[-0.5ex]
    \none &\quad\text{otherwise}
    \end{cases}
\end{align*}
\end{spreadlines}
\vspace{-1.0em}
\begin{align*}
    \opterm{P_1} \sqcup \opterm{P_2} \triangleq\ \none \quad\text{otherwise}
\end{align*}
\vspace{-1.5em}
\caption{The Merge Operation for Process Terms}
\label{merge:proc:fig}
\end{subfigure}
\begin{subfigure}{1.0\textwidth}
\begin{spreadlines}{0pt}
\begin{align*}
    &\some{\condif e_1 \condthen P_1 \condelse P_2} \sqcup \some{\condif e_2 \condthen Q_1 \condelse Q_2} \;\triangleq\\
    &\begin{cases}
    \;\some{\condif e_1 \condthen R_1 \condelse R_2} \quad &\text{if $e_1 = e_2$ and $\some{P_1} \sqcup \some{Q_1} \eqw \some{R_1}$ and $\some{P_2} \sqcup \some{Q_2} \eqw \some{R_2}$} \\[-0.5ex]
    \;\none &\text{otherwise}
    \end{cases}
\end{align*}
\end{spreadlines}
\vspace{-1.0em}
\begin{spreadlines}{0pt}
\begin{gather*}
    \some{P_1 \choice P_2} \sqcup \some{Q_1 \choice Q_2} \;\triangleq
    \begin{cases}
    \;\some{R_1 \choice R_2} \quad &\text{if $\some{P_1} \sqcup \some{Q_1} \eqw \some{R_1}$ and $\some{P_2} \sqcup \some{Q_2} \eqw \some{R_2}$} \\[-0.5ex]
    \;\none &\text{otherwise}
    \end{cases}\\[0.2em]
    \some{I} \sqcup \some{I} \triangleq\ \some{I} \quad\text{if $I$ is not a conditional or a choice} \quad\quad\quad\quad
    \opterm{I_1} \sqcup \opterm{I_2} \triangleq\ \none \quad\text{otherwise}
\end{gather*}
\end{spreadlines}
\vspace{-1.5em}
\caption{The Merge Operation for Process Instructions}
\label{merge:instr:fig}
\vspace{0.5em}
\end{subfigure}
\caption{The Merge Operation for Process Terms and Process Instructions}
\label{merge:fig}
\vspace{-1.0em}
\end{figure}

With the join-semilattice construction, we derive that $(\mbrs)$ is a \emph{partial order} on $\opt{Process}$ with properties: reflexivity, antisymmetry, and transitivity (cf. proposition~\ref{app:merge:mbrs:prop}). 
We establish a \emph{lax interchange law} (cf. \cref{merge:seqcomp:mbrs}) connecting merge $(\sqcup)$ and sequential composition $(\fatsemi)$ under $(\mbrs)$. We lift $(\fatsemi)$ to $\opt{Process}$, returning $\some{P \fatsemi Q}$ for $\some{P}$ and $\some{Q}$, and $\none$ if either argument is $\none$.
\begin{proposition}[Lax Interchange Law] \label[proposition]{merge:seqcomp:mbrs}
For any terms $\opterm{P_1}$, $\opterm{P_2}$, $\opterm{Q_1}$, and $\opterm{Q_2}$ in the set \emph{$\opt{Process}$}, the following relation holds:
\[
(\opterm{P_1} \sqcup \opterm{P_2}) \fatsemi (\opterm{Q_1} \sqcup \opterm{Q_2}) \mbrs (\opterm{P_1} \fatsemi \opterm{Q_1}) \sqcup (\opterm{P_2} \fatsemi \opterm{Q_2})
\]
\end{proposition} 
In \cref{merge:mbrs:net:def}, we define the more branches relation on networks by lifting the relation on $\opt{Process}$ pointwise to $\setof{Net}$.
\begin{definition}[$\,\mbrs$ on \emph{\setof{Net}}]\label{merge:mbrs:net:def}
    For any $N$ and $M$ in $\setof{Net}$, $N \mbrs M$ iff for any process $\pid{p}$, $\some{N (\pid{p})} \mbrsw \some{M (\pid{p})}$.
\end{definition}
We show that the relation $(\mbrs)$ is a \emph{simulation relation}, if $N \mbrs M$ and the smaller network $M$ performs a transition to $M'$, then $N$ can perform the same transition to some $N'$ such that $N' \mbrs M'$, as formalised in~\cref{merge:mbrs:net:transition}.
\begin{theorem}[Simulation under $\mbrs$]\label{merge:mbrs:net:transition}
    For any networks $N$ and $M$, if $N \mbrs M$ and $\langle M, \cstore, \mst, \mathscr{P} \rangle \xrightarrow[]{\mu} \langle M', \cstore', \mst', \mathscr{P} \rangle$, then there exists an $N'$ such that $\langle N, \cstore, \mst, \mathscr{P} \rangle \xrightarrow[]{\mu} \langle N', \cstore', \mst', \mathscr{P} \rangle$ and $N' \mbrs M'$.
\end{theorem}
\begin{proof}
   By structural induction on the derivation of the transition.
\end{proof}

%% file: appendix/coh.tex
\section{Coherence}
\label[appendix]{app:coherence}
We define \emph{coherence} between a choreography and choreographic procedure definitions for handling recursion correctly. It ensures the consistency of projection for partially entered procedure calls.
\begin{figure}[b]
\begin{mathpar}
\inferrule*[right={\scriptsize coh-end}]{ }{\mathscr{C} \propto \cnil}

\inferrule*[right={\scriptsize coh-local}]{\mathscr{C} \propto C}{\mathscr{C} \propto \pid{p}.x \metaDef e \seq C}

\inferrule*[right={\scriptsize coh-com}]{\mathscr{C} \propto C}{\mathscr{C} \propto \gencom \seq C}

\inferrule*[right={\scriptsize coh-sel}]{\mathscr{C} \propto C}{\mathscr{C} \propto \gensel \seq C}

\mathshade{
\inferrule*[right={\scriptsize coh-rtcom}]{\mathscr{C} \propto C}{\mathscr{C} \propto \genrtcom \seq C}
}

\mathshade{
\inferrule*[right={\scriptsize coh-rtsel}]{\mathscr{C} \propto C}{\mathscr{C} \propto \genrtsel \seq C}
}

\inferrule*[right={\scriptsize coh-cond}]{\mathscr{C} \propto C_1 \\ \mathscr{C} \propto C_2\\ \mathscr{C} \propto C}{\mathscr{C} \propto \condif \pid{p}.e\condthen C_1\condelse C_2 \seq C}

\inferrule*[right={\scriptsize coh-choice}]{\mathscr{C} \propto C_1 \\ \mathscr{C} \propto C_2 \\ \mathscr{C} \propto C}{\mathscr{C} \propto  C_1 \choice_\pid{p} C_2 \seq C}
\\
\inferrule*[right={\scriptsize coh-call}]{\mathscr{C} \propto C}{\mathscr{C} \propto X(\pids{p}) \seq C}
\qquad
\inferrule*[right={\scriptsize coh-rtcall},sep=0.7em]{
    \mathscr{C} \propto C \\ \mathscr{C} \propto C' \\ X(\pids{s}) = C'' \in \mathscr{C} \\\\
    |\pids{p}| = |\pids{s}| \\ \proj{C''[\pids{p}/\pids{s}] \fatsemi C'}{q} \eqr\some{P} \\ \proj{C}{q} \eqr\some{Q} \\ \some{P} \mbrsw \some{Q}}
{\mathscr{C} \propto \pid{q} : X(\pids{p}).C' \seq C}
\end{mathpar}
\caption{Coherence of Choreographies}
\label{epp:coherence:fig}
\end{figure}

%% file: appendix/algebraic-properties.tex
\section{Algebraic Properties of Operators}
\label[appendix]{app:algebraic-properties}
\begin{proposition} [Sequential Composition of Choreographies Forms a Monoid]\label{app:chor:seqcomp:prop}
\begin{align*}
    \cnil \fatsemi C &= C \quad\quad C \fatsemi \cnil = C &\text{(Identity element)}\\
    (C_1 \fatsemi C_2) \fatsemi C_3 &= C_1 \fatsemi (C_2 \fatsemi C_3) &\text{(Associativity)}
\end{align*}
\end{proposition}
\begin{proof}
By structural induction on the choreography.
\end{proof}

\begin{proposition} [Parallel Composition of Networks Forms a Monoid]\label[proposition]{app:proc:parcomp:prop}
\begin{align*}
    \nnil \parcomp N &= N \quad\quad N \parcomp \nnil = N &\text{(Identity element)}\\
    N_1 \parcomp (N_2 \parcomp N_3) &= (N_1 \parcomp N_2) \parcomp N_3 &\text{(Associativity)}
\end{align*}
\end{proposition}
\begin{proof}
By unfolding the definition of parallel composition and applying function extensionality.
\end{proof}

\begin{proposition} [Sequential Composition of Process Terms Forms a Monoid]\label{app:proc:seqcomp:prop}
\begin{align*}
    \pnil \fatsemi P &= P \quad\quad P \fatsemi \pnil = P &\text{(Identity element)}\\
    (P_1 \fatsemi P_2) \fatsemi P_3 &= P_1 \fatsemi (P_2 \fatsemi P_3) &\text{(Associativity)}
\end{align*}
\end{proposition}
\begin{proof}
By structural induction on the process term.
\end{proof}

\begin{proposition} [Sequential Composition of Networks Forms a Monoid]\label{app:net:seqcomp:prop}
\begin{align*}
    \nnil \fatsemi N &= N \quad\quad N \fatsemi \nnil = N &\text{(Identity element)}\\
    (N_1 \fatsemi N_2) \fatsemi N_3 &= N_1 \fatsemi (N_2 \fatsemi N_3) &\text{(Associativity)}
\end{align*}
\end{proposition}
\begin{proof}
By pointwise lifting from proposition~\ref{app:proc:seqcomp:prop}.
\end{proof}

\begin{proposition} [Join-Semilattice Properties of $(\opt{Process}, \sqcup)$]\label{app:merge:semilattice:prop}
\begin{align*}
    \opterm{P} \sqcup \opterm{P} &= \opterm{P} \quad&\text{(Idempotency)}\\
    \opterm{P} \sqcup \opterm{Q} &= \opterm{Q} \sqcup \opterm{P} \quad&\text{(Commutativity)}\\
    (\opterm{P} \sqcup \opterm{Q}) \sqcup \opterm{R} &= \opterm{P} \sqcup (\opterm{Q} \sqcup \opterm{R}) \quad&\text{(Associativity)}
\end{align*}
\end{proposition}
\begin{proof}
By structural induction on the process term.
\end{proof}

\begin{proposition} [The Relation $\mbrs$ is a Partial Order]\label{app:merge:mbrs:prop}
\begin{align*}
    \opterm{P} \mbrs \opterm{P} &\quad \quad&\text{(Reflexivity)}\\
    \opterm{P} \mbrs \opterm{Q} \wedge \opterm{Q} \mbrs \opterm{P} &\Rightarrow  \opterm{P} = \opterm{Q} \quad&\text{(Antisymmetry)}\\
    \opterm{P} \mbrs \opterm{Q} \wedge \opterm{Q} \mbrs \opterm{R} &\Rightarrow  \opterm{P} \mbrs \opterm{R} \quad&\text{(Transitivity)}
\end{align*}
\end{proposition}
\begin{proof}
    By the definition of the relation $\mbrs$ and join-semilattice properties of $(\opt{Process}, \sqcup)$.
\end{proof}

\begin{proposition} [Sequential Composition of Processes Distributes over Merge]\label{app:merge:seqcomp:prop}
\begin{align*}
   \opterm{P} \fatsemi (\opterm{Q} \sqcup \opterm{R}) = (\opterm{P} \fatsemi \opterm{Q}) \sqcup (\opterm{P} \fatsemi \opterm{R}) &\quad \quad&\text{(Left Distributivity)}\\
  (\opterm{P} \sqcup \opterm{Q}) \fatsemi \opterm{R} = (\opterm{P} \fatsemi \opterm{R}) \sqcup (\opterm{Q} \fatsemi \opterm{R}) &\quad \quad&\text{(Right Distributivity)}
\end{align*}
\end{proposition}
\begin{proof}
By structural induction on the process term.
\end{proof}

%% file: appendix/local-semantics.tex
\section{Localised Process Language Semantics}
\label[appendix]{app:local:semantics}
Because the asynchronous model does not require communication synchronisation, we can define a localised semantics for the process language. A transition by an active process has the form
$\langle P, \cstore, \mst, \mathscr{P} \rangle \xrightarrow[]{\mu}_{\pid{p}} \langle P', \cstore', \mst', \mathscr{P}\rangle$, where $\langle P, \cstore, \mst, \mathscr{P} \rangle$ is a process language configuration.
Formally, we define the localised semantics in \cref{local:semantics:fig}.

\begin{figure}
\vspace{-1.0em}
\begin{mathpar}
    \inferrule*[right={\scriptsize assign}]
    {\cstore(\pid{p}) \vdash e \downarrow v}
    {\langle x \metaDef e \seq P, \cstore, \mst, \mathscr{P} \rangle
     \xrightarrow{\tlassign{p}}_{\pid{p}}
     \langle P, \cstore[\pid{p}.x \mapsto v], \mst, \mathscr{P} \rangle}

    \inferrule*[right={\scriptsize send-val}]
    {\cstore(\pid{p}) \vdash e \downarrow v \\ \pid{p} \neq \pid{q} \\ \mst(\pid{q}, \pid{p}) = \vec{m}}
    {\langle \psend{q}{e} \seq P, \cstore, \mst, \mathscr{P} \rangle
     \xrightarrow{\tlcoms{p}{v}{q}}_{\pid{p}}
     \langle P, \cstore, \mst[(\pid{q}, \pid{p}) \mapsto \vec{m} \cons v], \mathscr{P} \rangle}

    \inferrule*[right={\scriptsize recv-val}]
    {\pid{p} \neq \pid{q} \\ \mst(\pid{q}, \pid{p}) = v \cons \vec{m}}
    {\langle \precv{p}{x} \seq Q, \cstore, \mst, \mathscr{P} \rangle
     \xrightarrow{\tlcomr{p}{v}{q}}_{\pid{q}}
     \langle Q, \cstore[\pid{q}.x \mapsto v], \mst[(\pid{q}, \pid{p}) \mapsto \vec{m}], \mathscr{P} \rangle}

    \inferrule*[right={\scriptsize send-sel}]
    {\pid{p} \neq \pid{q} \\ \mst(\pid{q}, \pid{p}) = \vec{m}}
    {\langle \psel{q} \seq P, \cstore, \mst, \mathscr{P} \rangle
     \xrightarrow{\tlsels{p}{q}{$\albl$}}_{\pid{p}}
     \langle P, \cstore, \mst[(\pid{q}, \pid{p}) \mapsto \vec{m} \cons \albl], \mathscr{P} \rangle}

    \inferrule*[right={\scriptsize recv-sel}]
    {j \in \idx{I} \\ \pid{p} \neq \pid{q} \\ \mst(\pid{q}, \pid{p}) = \albl_j \cons \vec{m}}
    {\langle \precvl{p}, \cstore, \mst, \mathscr{P} \rangle
     \xrightarrow{\tlselr{p}{q}{$\albl_j$}}_{\pid{q}}
     \langle P_j, \cstore, \mst[(\pid{q}, \pid{p}) \mapsto \vec{m}], \mathscr{P} \rangle}

    \inferrule*[right={\scriptsize cond-then}]
    {\cstore(\pid{p}) \vdash e \downarrow \mathit{true}}
    {\langle (\condif e \condthen P_1 \condelse P_2) \seq Q, \cstore, \mst, \mathscr{P} \rangle
     \xrightarrow{\tlthen{p}}_{\pid{p}}
     \langle P_1 \fatsemi Q, \cstore, \mst, \mathscr{P} \rangle}

    \inferrule*[right={\scriptsize cond-else}]
    {\cstore(\pid{p}) \vdash e \downarrow v \\ v \neq \mathit{true}}
    {\langle (\condif e \condthen P_1 \condelse P_2) \seq Q, \cstore, \mst, \mathscr{P} \rangle
     \xrightarrow{\tlelse{p}}_{\pid{p}}
     \langle P_2 \fatsemi Q, \cstore, \mst, \mathscr{P} \rangle}

    \inferrule*[right={\scriptsize call}]
    {X(\pids{r}) = Q \in \mathscr{P}}
    {\langle X(\pids{q}) \seq P, \cstore, \mst, \mathscr{P} \rangle
     \xrightarrow{\tlcall{p}}_{\pid{p}}
     \langle Q[\pids{q}/\pids{r}] \fatsemi P, \cstore, \mst, \mathscr{P} \rangle}

    \inferrule*[right={\scriptsize choice-l}]
    {\langle P, \cstore, \mst, \mathscr{P} \rangle \xrightarrow{\mu}_{\pid{p}}
     \langle P', \cstore', \mst', \mathscr{P} \rangle \\ \pid{p} \in \pn(\mu)}
    {\langle (P \choice Q) \seq R, \cstore, \mst, \mathscr{P} \rangle
     \xrightarrow{\tlleft{p}{\mu}}_{\pid{p}}
     \langle P' \fatsemi R, \cstore', \mst', \mathscr{P} \rangle}

    \inferrule*[right={\scriptsize choice-r}]
    {\langle Q, \cstore, \mst, \mathscr{P} \rangle \xrightarrow{\mu}_{\pid{p}}
     \langle Q', \cstore', \mst', \mathscr{P} \rangle \\ \pid{p} \in \pn(\mu)}
    {\langle (P \choice Q) \seq R, \cstore, \mst, \mathscr{P} \rangle
     \xrightarrow{\tlright{p}{\mu}}_{\pid{p}}
     \langle Q' \fatsemi R, \cstore', \mst', \mathscr{P} \rangle}

    \inferrule*[right={\scriptsize choice-skip}]
    { }
    {\langle (\pnil \choice \pnil) \seq P, \cstore, \mst, \mathscr{P} \rangle
     \xrightarrow{\tlskip{p}}_{\pid{p}}
     \langle P, \cstore, \mst, \mathscr{P} \rangle}
\end{mathpar}
\caption{Localised Semantics for the Process Language in \async}
\label{local:semantics:fig}
\vspace{-1.0em}
\end{figure}

\begin{lemma}[Local Transitions of Atomic Networks]\label[lemma]{net:local:to-network}
    Given a process name $\pid{p}$ and a transition label $\mu$, if $
    \langle P, \cstore, \mst, \mathscr{P} \rangle\xrightarrow{\mu}_{\pid{p}}\langle P', \cstore', \mst', \mathscr{P} \rangle$, then
    $\langle \pid{p}[P], \cstore, \mst, \mathscr{P} \rangle \xrightarrow{\mu}\langle \pid{p}[P'], \cstore', \mst', \mathscr{P} \rangle$.
\end{lemma}

\begin{lemma}[Lifting Local Transitions to Networks]\label[lemma]{net:local:of-local}
    Given networks $N$ and $N'$, a process name $\pid{p}$, and a transition
    label $\mu$, if $N(\pid{q}) = N'(\pid{q})$ for every $\pid{q} \neq \pid{p}$
    and $\langle N(\pid{p}), \cstore, \mst, \mathscr{P} \rangle \xrightarrow{\mu}_{\pid{p}} \langle N'(\pid{p}), \cstore', \mst', \mathscr{P} \rangle$,
    then $\langle N, \cstore, \mst, \mathscr{P} \rangle \xrightarrow{\mu} \langle N', \cstore', \mst', \mathscr{P} \rangle$.
\end{lemma}

\begin{lemma}[Localisation of Network Transitions]\label[lemma]{net:local:to-local}
    For a network configuration $\langle N, \cstore, \mst, \mathscr{P} \rangle$
    and a transition label $\mu$, if $\langle N, \cstore, \mst, \mathscr{P} \rangle \xrightarrow{\mu} \langle N', \cstore', \mst', \mathscr{P} \rangle$,
    then there exists a process name $\pid{p}$ such that
    $\pn(\mu) = \{\pid{p}\}$ and $\langle N(\pid{p}), \cstore, \mst, \mathscr{P} \rangle \xrightarrow{\mu}_{\pid{p}} \langle N'(\pid{p}), \cstore', \mst', \mathscr{P} \rangle$.
\end{lemma}

%% file: appendix/queue-transition.tex
\section{Localised Transitions for the Messaging State}
\label[appendix]{app:mst:semantics}

The transition $\mst \xrightarrow{\mu} \mst'$ of a messaging state with a transition label $\mu$, defined in \cref{mst:semantics:fig}.

\begin{figure}[htbp]
\vspace{-1.0em}
\begin{mathpar}
    \inferrule*[right={\scriptsize send-val}]
    { \mst(\pid{q},\pid{p}) = \vec{m} }
    {\mst \xrightarrow{\tlcoms{p}{v}{q}}
     \mst[(\pid{q},\pid{p}) \mapsto \vec{m} \cons v]}

    \inferrule*[right={\scriptsize recv-val}]
    {\mst(\pid{q},\pid{p}) = v \cons \vec{m}}
    {\mst \xrightarrow{\tlcomr{p}{v}{q}}
     \mst[(\pid{q},\pid{p}) \mapsto \vec{m}]}
    \\

    \inferrule*[right={\scriptsize send-sel}]
    { \mst(\pid{q},\pid{p}) = \vec{m} }
    {\mst \xrightarrow{\tlsels{p}{q}{l}}
     \mst[(\pid{q},\pid{p}) \mapsto \vec{m} \cons \albl]}

    \inferrule*[right={\scriptsize recv-sel}]
    {\mst(\pid{q},\pid{p}) = \albl \cons \vec{m}}
    {\mst \xrightarrow{\tlselr{p}{q}{l}}
     \mst[(\pid{q},\pid{p}) \mapsto \vec{m}]}
    \\

    \inferrule*[right={\scriptsize assign}]
    { }
    {\mst \xrightarrow{\tlassign{p}} \mst}

    \inferrule*[right={\scriptsize cond-then}]
    { }
    {\mst \xrightarrow{\tlthen{p}} \mst}

    \inferrule*[right={\scriptsize cond-else}]
    { }
    {\mst \xrightarrow{\tlelse{p}} \mst}

    \inferrule*[right={\scriptsize call}]
    { }
    {\mst \xrightarrow{\tlcall{p}} \mst}
    \\

    \inferrule*[right={\scriptsize choice-skip}]
    { }
    {\mst \xrightarrow{\tlskip{p}} \mst}

    \inferrule*[right={\scriptsize choice-l}]
    {\mst \xrightarrow{\mu} \mst'}
    {\mst \xrightarrow{\tlleft{p}{\mu}} \mst'}

    \inferrule*[right={\scriptsize choice-r}]
    {\mst \xrightarrow{\mu} \mst'}
    {\mst \xrightarrow{\tlright{p}{\mu}} \mst'}
\end{mathpar}
\caption{Messaging State Transitions in \async}
\label{mst:semantics:fig}
\vspace{-1.0em}
\end{figure}

\begin{lemma}[Removal of a Queue Head at an Uninvolved Receiver]\label[lemma]{mst:semantics:remove-passive-head}
    Given a messaging state transition $\mst \xrightarrow{\mu} \mst'$ and
    process names $\pid{p}$ and $\pid{q}$, if $\pid{q} \notin \pn(\mu)$
    and $\mst(\pid{q},\pid{p}) = \msg \cons \vec{m}$, then there exists
    a queue $\vec{m}'$ such that $\mst'(\pid{q},\pid{p}) = \msg \cons \vec{m}'$ and
    $\mst[(\pid{q},\pid{p}) \mapsto \vec{m}]
    \xrightarrow{\mu}
    \mst'[(\pid{q},\pid{p}) \mapsto \vec{m}']$.
\end{lemma}
\begin{proof}
    By structural induction on the derivation of $\mst \xrightarrow{\mu} \mst'$.
\end{proof}

\begin{lemma}[Isolating Messaging State Transitions]\label[lemma]{net:semantics:queue-step}
    If $\langle N, \cstore, \mst, \mathscr{P} \rangle
    \xrightarrow{\mu}
    \langle N', \cstore', \mst', \mathscr{P} \rangle$,
    then $\mst \xrightarrow{\mu} \mst'$.
\end{lemma}
\begin{proof}
    By structural induction on the derivation of the network transition.
\end{proof}

\begin{lemma}[Replacement of Messaging States in Choreographic Transitions]\label[lemma]{chor:semantics:queue-congr}
    Given a choreographic transition
    $\langle C, \cstore, \mst, \mathscr{C} \rangle \xrightarrow{\mu}
    \langle C', \cstore', \mst', \mathscr{C} \rangle$ and a messaging state
    transition $\mst_0 \xrightarrow{\mu} \mst_0'$, the choreography can perform
    $\langle C, \cstore, \mst_0, \mathscr{C} \rangle
    \xrightarrow{\mu}
    \langle C', \cstore', \mst_0', \mathscr{C} \rangle$.
\end{lemma}
\begin{proof}
    By structural induction on the derivation of the choreographic transition.
\end{proof}

%% file: appendix/finfun.tex
\section{Finite Functions and Finite Networks}
\label[appendix]{app:finfun}
\subsection{Finite Function}
\label[appendix]{app:finfun:finfun}
In our formalisation, a \emph{finite function} is a function with finite support. We write it as $\hat f : A \finfun B$.
For mechanisation, Mathlib~\cite{mathlib2020} provides a corresponding structure \code{Finsupp}, but some of its operations are noncomputable to avoid decidability requirements. To support computation, we introduce \code{FinFun}, pairing a function $f : A \to B$ with a proof of finite support, together with APIs that preserve this invariant.
Both \code{FinFun} and its APIs are included in CSLib, the Lean Computer Science Library.

\begin{definition}[Lifting Binary Operations to Finite Functions]
\label[definition]{app:finfun:mapBin}
Given a binary operation $\op$ defined on the set $\optg{B}$,
we lift $\op$ to finite functions $\hat{f}$ and $\hat{g}$
of the type $A \finfun B$:
\begin{align*}
    \fname{mapBin} \;\hat{f}\; \hat{g}\; \op \triangleq
    \begin{cases}
    \some{a \mapsto b_a} &\!\!\!\!\text{if $\support(\hat{f}) = \support(\hat{g})$ and for each $a \in \support(\hat{f}), \,\hat{f}(a) \op \hat{g}(a) =\,\some{b_a} $} \\[-0.5ex]
    \none &\!\!\!\!\text{otherwise}
    \end{cases}
\end{align*}
\end{definition}

Using $\fname{mapBin}$, we define the merge operation for
finite networks:
\[
\hat{N} \sqcup \hat{M} \triangleq
\fname{mapBin}\; \hat{N}\; \hat{M}\; \sqcup,
\]
where $(\sqcup)$ on the right-hand side is the merge operation
on $\opt{Process}$.

\subsection{Finite Network}
\label[appendix]{app:finfun:finnet}
EPP produces networks of finite support. \Cref{proc:finnet:def} shows such a \emph{finite network} ($\hat{N}$).
\begin{definition}[Finite Network]\label[definition]{proc:finnet:def}
A \emph{finite network} $\hat{N}$ is a \emph{finite function} (introduced in \cref{sec:finfun}) from process names to process terms, i.e.,
$\hat{N} : \textbf{PName} \to_f \textbf{Process}$.
\end{definition}

%% file: appendix/process-ops.tex
\section{Process Removal and Network Restriction}
\label[appendix]{app:net:rm-res}
We have the \emph{process removal} and the \emph{network restriction} operations (cf. \cref{proc:network:removal-restriction:def}).
The process removal operation $N \setminus \pid{p}$ updates $N$ by setting $N(\pid{p}) = \pnil$. Process removal is commutative and associative. We can generalise it to finite sets of processes, written as $N \setminus \{\pids{p}\}$.
The network restriction operation is the opposite, which extracts a subnetwork containing only a finite set of processes $\{\pids{p}\}$ from a given network $N$.
The resulting subnetwork is a finite network written as $N \restrict \{\pids{p}\}$.
\begin{figure}[b]
\vspace{-1.0em}
\begin{spreadlines}{0pt}
\begin{align*}
\begin{aligned}[c]
    (N \setminus \pid{p})(\pid{q}) &\triangleq 
    \begin{cases}
    \ \pnil &\text{if $\pid{p} = \pid{q}$}\\[-0.5ex]
    \ N(\pid{q}) &\text{otherwise}
    \end{cases}
\end{aligned}
\quad\quad
\begin{aligned}[c]
    (N \restrict \{\pids{p}\})(\pid{p}) &\triangleq 
    \begin{cases}
    \ N(\pid{p}) &\text{if $\pid{p} \in \{\pids{p}\}$}\\[-0.5ex]
    \ \pnil &\text{otherwise}
    \end{cases}
\end{aligned}
\end{align*}
\end{spreadlines}
\caption{Process Removal and Network Restriction}
\label{proc:network:removal-restriction:def}
\vspace{-1.0em}
\end{figure}

%% file: appendix/semantic-properties.tex
\section{Properties of the Semantics of Choreographies}
\label[appendix]{app:chor:lts:props}
If a choreography can perform a transition with label $\mu$, then the choreography must contains all the processes in the transition label $\mu$, which is shown in theorem~\ref{chor:semantics:involved-processes}.
\begin{theorem}[Transitions Involve Only Processes in the Choreography]
\label{chor:semantics:involved-processes}
For any choreographic configuration $\langle C, \cstore, \mathscr{C} \rangle$ and transition label $\mu$, if $\langle C, \cstore, \mathscr{C} \rangle \xrightarrow[]{\mu} \langle C', \cstore', \mathscr{C} \rangle$, then \emph{$\pn(\mu) \subseteq \pn(C)$}.
\end{theorem}
\begin{proof}
    By structural induction on the derivation of the transition.
\end{proof}

A choreographic transition may change the projection onto a process even when that process is not involved in the transition. The following theorem states that the projection onto that process before the transition offers every branch that its projection after the transition offers,
\begin{theorem}[Projection of Uninvolved Processes]\label{epp:projection:uninvolved-processes} Given a well-formed choreographic configuration $\langle C,\cstore,\mathscr{C}\rangle$, if $\langle C,\cstore,\mathscr{C}\rangle \xrightarrow{\mu} \langle C',\cstore',\mathscr{C}\rangle$, then for any process \emph{$\pid{p}\notin\pn(\mu)$}, $\proj{C}{p}\mbrs\proj{C'}{p}$.
\end{theorem}
\begin{proof} By structural induction on the derivation of the transition $\langle C,\cstore,\mathscr{C}\rangle \xrightarrow{\mu} \langle C',\cstore',\mathscr{C}\rangle$.
\end{proof}

\section{Properties of the Semantics of Networks}
\label[appendix]{app:net:lts:props}
As in the choreographic case, the processes involved in a network transition are exactly those named in its label, shown in theorem~\ref{net:semantics:involved-processes}.
\begin{theorem}[Transitions Involve Only Running Processes]
    \label{net:semantics:involved-processes}
    For any network configuration $\langle N, \cstore, \mathscr{P} \rangle$ and transition label $\mu$, if $\langle N, \cstore, \mathscr{P} \rangle \xrightarrow[]{\mu} \langle N', \cstore', \mathscr{P} \rangle$, then \emph{$\pn(\mu)\subseteq \support(N)$}.
\end{theorem}
\begin{proof}
    By structural induction on the derivation of the transition.
\end{proof}
Based on this property, we establish two further properties of network transitions involving network restriction and process removal.
If a network can perform a transition, then its subnetwork containing only the processes named in the transition label can perform the same transition, shown in theorem~\ref{net:semantics:restriction}.

\begin{theorem}[Restriction to Involved Processes Preserves Transitions]
    \label{net:semantics:restriction}
    For any network configuration $\langle N, \cstore, \mathscr{P} \rangle$ and transition label $\mu$, if $\langle N, \cstore, \mathscr{P} \rangle \xrightarrow[]{\mu} \langle N', \cstore', \mathscr{P} \rangle$, then \emph{$\langle N\restrict \pn(\mu), \cstore, \mathscr{P} \rangle \xrightarrow[]{\mu} \langle N'\restrict \pn(\mu), \cstore', \mathscr{P} \rangle$}.
\end{theorem}
\begin{proof}
    By structural induction on the derivation of the transition.
\end{proof}
Similarly, removing processes that the transition label does not mention preserves the transition, shown in theorem~\ref{net:semantics:proc-removal}.
\begin{theorem}[Removal of Uninvolved Processes Preserves Transitions]
    \label{net:semantics:proc-removal}
    For any network configuration $\langle N, \cstore, \mathscr{P} \rangle$ and transition label $\mu$, if $\langle N, \cstore, \mathscr{P} \rangle \xrightarrow[]{\mu} \langle N', \cstore', \mathscr{P} \rangle$, then for any finite set of processes  such that \emph{$\{\pids{p}\} \disjoint \pn(\mu)$}, $\langle N\setminus \{\pids{p}\} , \cstore, \mathscr{P} \rangle \xrightarrow[]{\mu} \langle N'\setminus \{\pids{p}\}, \cstore', \mathscr{P} \rangle$.
\end{theorem}
\begin{proof}
    By structural induction on the derivation of the transition.
\end{proof}

\begin{theorem}[Lax Compatibility of Merge at the Active Process]\label{merge:net:transition}
    For any process terms $P$, $Q$, $P'$, and $Q'$ such that every receive-branch list in $P$ and $Q$ is strictly ordered by label and $\some{P} \sqcup \some{Q} \eqw \some{R}$,
    if $\langle P, \cstore, \mst, \mathscr{P} \rangle \xrightarrow{\mu}_{\pid{p}} \langle P', \cstore_1', \mst_1', \mathscr{P} \rangle$ and $\langle Q, \cstore, \mst, \mathscr{P} \rangle \xrightarrow{\mu}_{\pid{p}} \langle Q', \cstore_2', \mst_2', \mathscr{P} \rangle$,
    then there exists a process term $S$ such that $\langle R, \cstore, \mst, \mathscr{P} \rangle \xrightarrow{\mu}_{\pid{p}} \langle S, \cstore_1', \mst_1', \mathscr{P} \rangle$ and $\some{S} \;\mbrs (\some{P'} \sqcup \some{Q'})$.
\end{theorem}
\begin{proof}
    By induction on the local transition derivation for $P$, using \cref{merge:seqcomp:mbrs} when reattaching continuations.
\end{proof}

%% file: appendix/misc.tex
\section{Coherence}
\label[appendix]{app:epp:coherence}
We present the formal definition of coherence between choreographies and their corresponding sets of choreographic procedures in figure~\ref{epp:coherence:fig}.

\section{Multi-step Transitions for Choreographies and Networks}
\label[appendix]{app:multi}
The formal definitions of multi-step transitions for both choreographies and networks are shown in \cref{chor:semantics:multi-step} and \cref{net:semantics:multi-step}, respectively. 
\begin{figure}
\begin{subfigure}{1.0\textwidth}
\begin{mathpar}
    \inferrule*[right={\scriptsize refl}]{ }{\langle C, \cstore, \mst, \mathscr{C} \rangle \xtwoheadrightarrow{\,\epsilon\,} \langle C, \cstore, \mst, \mathscr{C} \rangle}

    \inferrule*[right={\scriptsize step-l}] {\langle C, \cstore, \mst, \mathscr{C} \rangle \xrightarrow[]{\mu} \langle C', \cstore', \mst', \mathscr{C} \rangle \\ \langle C', \cstore', \mst', \mathscr{C} \rangle\xtwoheadrightarrow{\,\tls\,}  \langle C'', \cstore'', \mst'', \mathscr{C} \rangle }{\langle C, \cstore, \mst, \mathscr{C} \rangle \xtwoheadrightarrow{\,\mu \,\cons\, \tls\,} \langle C'', \cstore'', \mst'', \mathscr{C} \rangle}
\end{mathpar}
\caption{Multi-Step Transitions for Choreographies}
\label{chor:semantics:multi-step}
\end{subfigure}
\begin{subfigure}{1.0\textwidth}
\begin{mathpar}
    \inferrule*[right={\scriptsize refl}]{ }{\langle N, \cstore, \mst, \mathscr{P} \rangle \xtwoheadrightarrow{\,\epsilon\,} \langle N, \cstore, \mst, \mathscr{P} \rangle}

    \inferrule*[right={\scriptsize step-l}] {\langle N, \cstore, \mst, \mathscr{P} \rangle \xrightarrow[]{\mu} \langle N', \cstore', \mst', \mathscr{P} \rangle \\ \langle N', \cstore', \mst', \mathscr{P} \rangle\xtwoheadrightarrow{\,\tls\,}  \langle N'', \cstore'', \mst'', \mathscr{P} \rangle }{\langle N, \cstore, \mst, \mathscr{P} \rangle \xtwoheadrightarrow{\,\mu \,\cons\, \tls\,} \langle N'', \cstore'', \mst'', \mathscr{P} \rangle}
\end{mathpar}
\caption{Multi-Step Transitions for Networks}
\label{net:semantics:multi-step}
\end{subfigure}
\caption{Multi-Step Transitions}
\label{app:semantics:multi-step}
\end{figure}
Both multi-step transition relations have two rules. The rule \textsc{refl} allows any configuration to take a multi-step transition to itself with an empty transition label. The rule \textsc{step-l} states that if a configuration takes a single-step transition to a second configuration with label $\mu$, and the second configuration takes a multi-step transition to a third configuration with label $\tls$, then the first configuration can take a multi-step transition to the third configuration with a label obtained by prepending $\mu$ to $\tls$. In our Lean~4 mechanisation, we instantiate CSLib's multi-step transition relation \code{MTr} for choreographies and networks.

\section{EPP Produces Communication Safe States}
\begin{lemma}[Communication Safety with More Branches]\label{epp:comm-safe-state:more-branches}
    If $M$ is communication safe at $\mst$, then $N$ is communication safe at $\mst$.
\end{lemma}
\begin{proof}
    By case analysis on process merge.
\end{proof}

\begin{theorem}[EPP Produces Communication-Safe States]\label{epp:comm-safe-state}
    If $\langle C,\cstore,\mst,\mathscr{C}\rangle$ is well-formed, $C\propto\mathscr{C}$, and $\epp{C}=\;\some{N}$, then $N$ is communication safe at $\mst$.
\end{theorem}
\begin{proof}
    By induction on $C$.
\end{proof}